\pdfoutput=1
\documentclass[twocolumn,
 showpacs,
 amsmath,amssymb,
prb,
twocolumn
]{revtex4-1}

\usepackage{graphicx}
\usepackage{dcolumn}
\usepackage{bm}
\usepackage[mathlines]{lineno}

\usepackage[backref=none]{hyperref} 
\usepackage[usenames,dvipsnames]{color}
\definecolor{MyDarkBlue}{rgb}{0,0.1,0.7}
\hypersetup{pdfborder={0 0 0},colorlinks,breaklinks=true,
  urlcolor={MyDarkBlue},citecolor={MyDarkBlue},linkcolor={MyDarkBlue}
}
\usepackage{booktabs}
\newcolumntype{C}{>{$}c<{$}}
\AtBeginDocument{
\heavyrulewidth=.08em
\lightrulewidth=.05em
\cmidrulewidth=.03em
\belowrulesep=.65ex
\belowbottomsep=0pt
\aboverulesep=.4ex
\abovetopsep=0pt
\cmidrulesep=\doublerulesep
\cmidrulekern=.5em
\defaultaddspace=.5em
}
\usepackage{longtable}
\usepackage{xtab}
\begin{document}


\title{First principles soft mode lattice dynamics of PbZr$_{0.5}$Ti$_{0.5}$O$_3$ and shortcomings of the virtual crystal approximation}

\author{Jack S. Baker$^{1, 2}$}
\author{David Bowler$^{1, 2}$}%

\affiliation{
$^1$London Centre for Nanotechnology, UCL, 17-19 Gordon St, London WC1H 0AH, UK \\ $^2$Department of Physics \& Astronomy, UCL, Gower St, London WC1E 6BT, UK}


\date{\today}

\begin{abstract}

A comparative study between PbTiO$_3$, PbZrO$_3$, and the solid solution PbZr$_{0.5}$Ti$_{0.5}$O$_3$ is performed on the soft mode lattice dynamics within the first Brillouin Zone. We consider the six unique B-site orderings for PbZr$_{0.5}$Ti$_{0.5}$O$_3$ representable within the 2$\times$2$\times$2 primitive perovskite supercell as well as the virtual crystal approximation (VCA) to extract the phonon dispersion relations of a high-symmetry cubic-constrained form using density functional perturbation theory. We find that the most unstable modes in the rock-salt ordered structure and the VCA, like pure PbZrO$_3$, are antiferrodistortive (AFD) whilst lower symmetry arrangements are dominated by $\Gamma$-point ferroelectric (FE) instabilities like pure PbTiO$_3$. Despite similarities in the phonon dispersion relations between the rock-salt ordered supercell and the VCA, the character of modes at high symmetry points are found to be different. In particular, the a$^{0}$a$^{0}$c$^{-}$ \& a$^{0}$a$^{0}$c$^{+}$ AFD instabilities of the rock-salt ordering are replaced with a$^{-}$b$^{-}$c$^{-}$ \& a$^{+}$b$^{+}$c$^{+}$ instabilities within the VCA. Such a rotation pattern is not seen in any of the supercell-based calculations thus serving as a quantitative example of the inability of the method to represent accurately local structural distortions. Single modes are found exhibiting dual order parameters. At the zone centre, some arrangements show mixed FE \& antipolar soft modes (due to Pb motion tansverse to the polar axis) and at long wavelengths all arrangements have soft modes of a mixed antipolar \& AFD character. These are described with direct analysis of the eigendisplacements.


\end{abstract}
\maketitle


\section{\label{intro:level1} Introduction}
The PbZr$_x$Ti$_{1-x}$O$_3$ (PZT) solid solution is the most abundantly used piezoelectric material. This is due to its giant electromechanical response and well developed, low cost synthesis \cite{jaffe2012piezoelectric, Oliveira2014}. Together, this has ensured the technological relevance of the material; well adapted for exploitation in ultrasonic transducers \cite{Izyumskaya2007, gururaja1985piezoelectric}, ceramic capacitors and actuators \cite{Chi2014}. More exotically, PZT has been proposed for use in potential piezoelectricity-induced room temperature superconductors where a supercurrent is induced along a metal/piezoelectric interface \cite{Kadin2017,SalvatorePais2019}. For these applications, it is most common to consider PZT at around $x\approx0.52$ \cite{Jin2003, Agar2014} in the region near the morphotropic phase boundary (MPB). This is a compositional boundary at the peak of the electromechanical response. This boundary exhibits complex lattice dynamics where a flat energy surface for polarization rotation exists between the FE tetragonal ($\mathbf{P}$ $\parallel$ [001]) and rhombohedral ($\mathbf{P}$ $\parallel$ [111]) phases via intermediate monoclinic phases \cite{Jin2003, Noheda1999, Catalan2011}.

It is useful to regard PZT as a randomly ordered isovalent B-site substituted compound in a matrix of either of the two phase diagram end members PbTiO$_3$ (PTO) or PbZrO$_3$ (PZO). The former is a prototypical FE with  $P4mm$ symmetry \cite{Nelmes1985} whilst the latter, though still topical \cite{Fthenakis2017,rabe2013antiferroelectricity,Tagantsev2013, Hlinka2014, Mani2015}, is considered an antiferroelectric (AFE) with $Pbam$ symmetry. These observations are supported using the soft mode theory of lattice dynamics by considering the symmetry (and energy) lowering distortions of a high-symmetry cubic phase as indicated by imaginary frequencies at certain wavevectors in the phonon spectrum \cite{RAMAN1940, Cochran1959, Cochran1960, anderson1960pw}. It is using this method that the modes responsible for the paraelectric to FE transition in PTO and paraelectric to AFE transition in PZO are identified as $\Gamma_4^-$ \cite{Nelmes1985,Sicron1994} and dual $\Sigma_2$ + R$_4^+$ (and to a lesser extent, S$_4$, R$_5^+$, X$_3^-$ and M$_5^-$) \cite{Tagantsev2013, Mani2015} respectively. Such a classification is not possible for a truly random alloy. Even for ordered PZT it proves much more difficult since the character and frequencies of the relevant modes may vary with Ti/Zr concentration as well as with the specific ordering of the B-site substitutions in the crystal lattice, for which, in a periodic crystal the number of permutations are infinite.\par

In order to study PZT near the MPB with first principles calculations, we consider two paths. Both paths impose fictitious symmetry when compared to the real random compound. The first is to explore the different permutations of Ti/Zr substitutions within a supercell of finite size. True morphotropic PZT requires simulation in a large supercell so $x=0.5$ is often chosen as a surrogate. This is the most common approach taken and has been successful in the calculation of structural \cite{Marton2011, Blok2011}, piezoelectric \cite{Wu2003, Kim2013} and electronic properties \cite{Grinberg2004}. Using this method, phonon disperison relations across a small area of the first Brillouin zone have also been calculated for [1:1] PZO/PTO superlattices \cite{Bungaro2002}. For (001) and (110) ordered structures, FE modes were isolated to Ti/Zr layers whilst the (111) ordered superlattice displays one mode behaviour with competing FE and AFD character. This study, however, was limited in scope by only considering modes at the zone centre. The second option is to use a mixed potential scheme such as the VCA. This approach, like the supercell method, predicts anomalous dynamical charges and with reasonable accuracy, the location of the MPB \cite{Bellaiche2000, Ramer2000, Liu2013} but is unable to accurately represent distortions to local structure. The extent to which this is true, however, is unknown thus a quantitative comparison based on the characteristics of the soft mode distortions would be valuable. This approach, however, does allow access to a wide range of Ti/Zr concentrations at a fraction of the computational cost of a large supercell calculation. \par

It is the aim of this work to provide a complete comparative study of the phonon dispersion relations in near-morphotropic PbZr$_{0.5}$Ti$_{0.5}$O$_3$ within density functional theory (DFT) using the VCA and supercell method complete with comparison to the end members PTO and PZO. We do so also with special consideration of longer wavelength modes often not considered. We compare the characters of soft modes by considering distortions at high symmetry points via eigendisplacement analysis and the projected phonon density of states (PDOS). Doing so gives access to displacement patterns and to the species specific character of all modes in the soft space. We select the $2 \times 2 \times 2$ supercell of the primitive perovskite unit for our simulations to coincide with measured mean cluster size distributions for Ti/Zr ordering in PZT \cite{Bell2006}. Such supercells have recently been used as local phases to build a complex multiphase model of the material able to predict the experimental pair distribution function to a high accuracy \cite{Bogdanov2016}. Such a supercell dimension is also important for theoretical studies since important competitive modes inlcuding Glazer-like \cite{Glazer1972, Glazer1975} AFD, FE and some AFE modes fold to the zone-centre. However, since our calculations are performed throughout the full first Brillouin zone, we are not limited to the zone centre and so we can identify competitive long wavelength order not usually considered in PZT.  We obtain the irreducible representations (irreps) of the soft mode distortions and identify their incipient order parameters which in the case of longer wavelength modes we find can impose dual order. By doing so, we provide further insight into the complex lattice dynamics occuring near the MPB. Further, it will provide a guide for future investigations detailing the consequences of using the supercell or VCA methods for future studies of PZT and heterostructures for which PZT is an ingredient. \par

The rest of this work is organised as follows. In section \ref{method:level1} we detail the theoretical methods for the calculations, including details for the calculation of the electronic ground-state, phonon dispersions and details for the specific implementation of the VCA. In section \ref{results:level2} we discuss the properties of the fully relaxed parent structures. Then, in section \ref{phonon:level2} we present the full phonon dispersion relations and PDOS along with a discussion and tabulation of the relevant soft modes and their frequencies. We begin first with a comparison between the end members PTO \& PZO. The other dispersions are then paired based on their similarity and discussed together with the exception of $Pm\bar{3}m$ ordered PZT supercell which is dedicated its own section. Modes important to the disussion are shown graphically. These results are then discussed more broadly and summarised in section \ref{summary:level1}.

\section{\label{method:level1} Theoretical method}

Calculations are performed using the implementation of DFT as present in the \texttt{ABINIT} code (\texttt{v8.10.2}) \cite{Gonze2016, Gonze2009}. We use scalar-relativistic, norm-conserving pseudopotentials generated by the \texttt{ONCVPSP} code (\texttt{v0.3}) \cite{hamann2013optimized} as made available on the \texttt{PseudoDojo} website \cite{van2018pseudodojo}. These potentials treat the Pb 5d$^{10}$6s$^2$6p$^6$, Ti 3s$^2$3p$^6$4s$^2$3d$^{10}$, Zr 4s$^2$4p$^6$5s$^2$5d$^{10}$ and O 2s$^2$2p$^6$ orbitals as valence. These pseudopotetial include partial core corrections. For the $2 \times 2 \times 2$ supercells, Brillouin zone integrals are performed with sums over $\Gamma$-centered $4 \times 4 \times 4$ Monkhorst-Pack \cite{monkhorst1976special} meshes. A plane-wave cutoff energy of 1088.46 eV (40 Ha) is employed to ensure the accuracy of our calculations. Exchange \& correlation effects are represented by the PBESol \cite{Perdew2008} functional as present in \texttt{Libxc} (\texttt{v3.0.0}) \cite{Marques2012}. This functional is known to produce high accuracy structural properties compared with experiment \cite{Zhang2017} justifying its use in a study of structural distortion. This method returns the paraelectric cubic $Pm\bar{3}m$ lattice constants of PTO and PZO as $a_{\text{PTO}}=3.918$ $\text{\AA}$ (-0.304\%) and $a_{\text{PZO}}=4.140$ $\text{\AA}$ (+0.242\%) where bracketed values are errors compared with experiment \cite{Mabud1979, Sawaguchi1953}. We make particular use of the linear response features in \texttt{ABINIT} for the calculation of phonon dispersions using density functional perturbation theory (DFPT) \cite{Gonze1997, Baroni2001}. Dynamical matrices are calculated on the $\mathbf{q}$-point mesh of the supercell calculation and dispersion is extracted using a fourier interpolation scheme between points on the $\mathbf{q}$-point mesh \cite{Gonze1997, Baroni2001}. Since the perovskite oxides are known to give rise to giant LO-TO splitting \cite{Zhong1994}, we require the non-analytic correction (NAC) at the $\Gamma$-point \cite{Gonze1997} to correct for the undefined nature of the long-range Coulomb interactions \cite{Henry1965}. This correction requires knowledge of the electronic dielectric tensor $\mathbf{\epsilon}^{\infty}$ and Born effective charges $\mathbf{Z}^*_i$ where $i$ labels each atomic site in the supercell. Both are obtained also using DFPT in response to a homogeneous electric field \cite{Gonze1997-2, Gonze1997}. \par

For calculations involving use of the VCA, we use the implementation in \texttt{ABINIT}. It is used to create an '\textit{alchemical}' virtual atom of Ti/Zr character by linearly mixing the pseudopotentials of the individual species.

\begin{equation}
  V_{\text{VCA}}^{\text{ps}} = xV_{\text{Zr}}^{\text{ps}} + (1-x)V_{\text{Ti}}^{\text{ps}}
  \label{VCA}
\end{equation}

This can be further broken down into local contributions and short-range non-local corrections \cite{Ghosez2000}. Phonon dispersion calculations using DFPT \textit{and} the VCA are currently not fully supported in the code so we instead use the (formally equivalent) finite displacement method (FDM) as implemented in the \texttt{Phonopy} code (\texttt{v2.1}) \cite{Togo2015} using a $4\times 4 \times 4$ supercell of the primitive perovskite unit and a displacement of 0.01 \AA. For this calculation, the virtual atom must take on the intermediate mass of Ti and Zr, equal to 69.55 AMU. The NAC is accounted for following the same method as used in the DFPT calculations. For means of validation, a comparison of the phonon dispersions for PTO \& PZO using both DFPT and the FDM are given in section 2 of the Supplemental Material\cite{note:SM}. \par

For section \ref{phonon:level2}, we treat the six unique B-site configurations of PbZr$_{0.5}$Ti$_{0.5}$O$_3$ within the $2\times 2\times 2$ supercell labelled with Roman numerals I:VI. These supercells are shown in FIG \ref{fig:ConfigsPZT}. Although PTO, PZO and VCA calculations are representable in the primitive perovskite cell, we still choose to use the $2\times 2\times 2$ supercell such that phonon dispersions are calculated along the same $\mathbf{q}$-path as for structures I:VI and share the same total number of phonon branches ($3 \times N_{\text{atom}}=120$). PZT supercells are constrained to be cubic with dimensions (2$a_{\text{Vg}}$, 2$a_{\text{Vg}}$, 2$a_{\text{Vg}}$) where $a_{\text{Vg}} = 4.029$ $\text{\AA}$, the lattice constant set by Vegard's law \cite{Vegard1921}. For $x=0.5$, this is a simple average of $a_{\text{PTO}}$ and $a_{\text{PZO}}$. This choice of lattice constant  favours no particular B-site ordering that may be biased in different experimental conditions. Further, structural data for high temperature cubic PZT is scarce since the technologically relevant large piezoelectric coefficients stem from the low temperature tetragonal/rhombohedral phases. Simulations for PTO and PZO are performed at their theoretical lattice constants. Before the phonon calculations, internal degrees of freedom are relaxed to a stringent force tolerance of $1\times 10^{-6}$ eV/$\text{\AA}$ to prevent soft modes forming from non-equilibrium vibrations. To further illuminate the mode characters, we also calculate the phonon PDOS for each structure. To do so, we calculate the dynamical matrix on a dense 49$\times$49$\times$49 grid of $\mathbf{q}$-points and integrate with the tetrahedron method \cite{Blchl1994}. \par


Throughout this work, we make use of group theoretical software. We use the programs \texttt{FINDSYM} (\texttt{v6.0}) \cite{Stokes2005} and \texttt{ISODISTORT} (\texttt{v6.5}) \cite{Campbell2006} as made available in the \texttt{ISOTROPY} software suite. We also make use of the web-based phonon spectrum visualisation tools made available by H. Miranda \cite{PhononWeb}.

\section{\label{results:level1} Results}
\subsection{\label{results:level2} Parent structures}

    \begin{figure}
       \includegraphics[width=\linewidth]{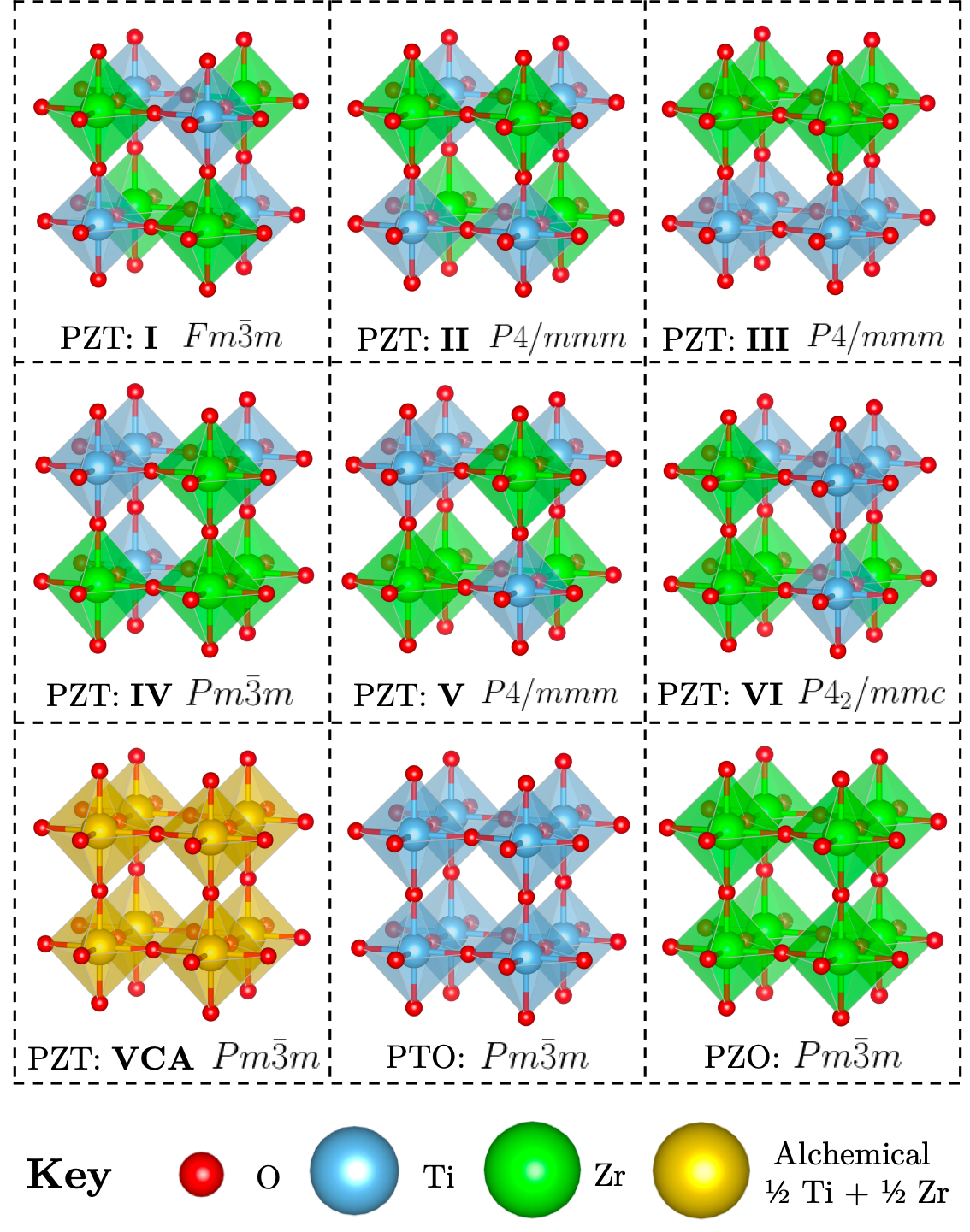}
        \caption{The structures used for the phonon dispersion calculations of section \ref{phonon:level2}. A-site Pb has been removed for clarity and BO$_6$ octahedral complexes have been coloured to match the B-site species. Supercell models (rows 1 and 2) are labelled with Roman numerals I:VI whilst the last row indicates the PZT-VCA supercell as well as the end members PTO and PZO. Each supercell is also assigned a crystalline space-group.}
        \label{fig:ConfigsPZT}
    \end{figure}

Table \ref{tab:ConfigPZT} details the structural and symmetry properties of the relaxed primitive cells. We find that a simple metric like the number of Wyckoff sites (and their deviation from the ideal perovskite sites) suggests which arrangements have comparable lattice dynamics. This is used  as a basis for for the discussion in section \ref{phonon:level2}. These primitive cells are then translated into the 2$\times$2$\times$2 supercell of the primitive PbBO$_3$ unit (B=Zr or B=Ti) and are shown in figure \ref{fig:ConfigsPZT}. These form a set of parent structures from which we later perform mode decompositional analysis. Table \ref{tab:BECDielecSigma} shows other important structural, dynamical and dielectric properties also important to the discussion in this section. \par 

PTO/PZO/VCA cells show the usual cubic $Pm\bar{3}m$ symmetry. These are joined by PZT I ($Fm\bar{3}m$) and IV ($Pm\bar{3}m$) which also support a cubic local minimum. The former adopts rock-salt-like ordering with continuous B-sites aligned along the [111] direction whilst the latter shows a separation of Ti and Zr sites into opposite corners of the supercell. As a consequence, these parents show isotropic behaviour in both the stress and high frequency dielectric tensor (table \ref{tab:BECDielecSigma}). This is in contrast to the other four PZT parents which are members of lower symmetry tetragonal spacegroups (even whilst constrained to $a_{\text{Vg}}$) thus showing anisotropic behaviour in these tensors about a single axis. It is typical behaviour across all of the PZT parents (bar the VCA) to compress areas of TiO$_6$ coordination making way for the larger ZrO$_6$ octahedra. When constrained to $a_{\text{vg}}$, PZT I is the most energetically stable configuration whilst III is the most unstable with an energy difference of 114 meV/PbBO$_3$ unit between them. Remarkably, if we perform a full cell shape and size relaxation, this energy difference marginally narrows to 111 meV/PbBO$_3$ showing the small contribution of strain energy to the non-polar phases of PZT. \par

Table \ref{tab:BECDielecSigma} indicates that at $a_{\text{vg}}$, PZT is held at a non-vanishing pressure. The VCA exhibits the largest $\sigma_{\text{RMS}}$ of 2.62 GPa whilst II and III show stronger uniaxial stress about the axes of compositional modulation indicating a proclivity for expansion in these directions. PZT I:VI shows remarkably similar $\mathbf{Z}^*$ and $\epsilon^{\infty}$ indicating that Ti/Zr cation ordering has little influence on these quantities. It is also notable that $\mathbf{Z}^*$ of PZT I:VI deviates only a small amount from the mean  $\mathbf{Z}^*$ of PZO and PTO.  The VCA shows good agreement with the supercell method for $\bar{Z}_{\text{Pb}}^*$ and $\bar{Z}^*_{\text{O}_{\perp}}$ but underestimates strongly the magnitudes of the alchemical $\bar{Z}_{\text{B}}^*$ and $\bar{Z}^*_{\text{O}{\parallel}}$. The VCA also features a strong discrepancy in $\mathbf{\epsilon}^{\infty}$ compared to both the mean and supercell approach. Although not tabulated, it should be noted that PZT II, IV, V and VI feature off-diagonal elements in the BEC tensor only for $\bar{Z}_{\text{Pb}}^*$. These components are small and do not exceed 0.34 electronic charges in magnitude but do vary in sign despite the positive nature of the Pb cation. It should also be noted that using a similar method, a previous study reports off-diagonal elements not of Pb, but of the O 4k site, always negative in sign \cite{Bungaro2002}.

\LTcapwidth=\columnwidth
\begin{longtable}{@{}cc@{}}
\caption{The space-groups, Wyckoff positions and \textit{primitive} supercell dimensions of the relaxed, cubic-constrained structures. All structures are representable in the $2\times 2 \times 2$ primitive perovskite supercell as illustrated in figure \ref{fig:ConfigsPZT}. Wyckoff positions are stated using the site multiplicity and Wyckoff letter as made standard by the Bilbao Crystallographic Server \cite{Aroyo2006}. Supercell dimensions are given in lengths of the mutually orthogonal axes a, b \& c with the exception of PZT I whose axes are at an angle $\alpha = \beta = \gamma = 60^{\circ}$ with the full form of the lattice vectors displayed.}
\label{tab:ConfigPZT}\\ \\
\multicolumn{2}{c}{\textbf{PTO/PZO/VCA ($Pm\bar{3}m$ $O_h^1$)}}                                                \\* \midrule \midrule
\endfirsthead
\multicolumn{2}{c}{$a=a_{\text{PTO}}$/$a_{\text{PZO}}$/$a_{\text{Vg}}$}                                                                                                                                                                                                                                                                             \\* \midrule
Pb 1b                                                                                                                                                                  & (1/2, 1/2, 1/2)                                                                                                                                                            \\
Ti/Zr/(1/2 Ti + 1/2 Zr) 1a                                                                                                                                             & (0, 0, 0)                                                                                                                                                                  \\
O 3d                                                                                                                                                                   & (1/2, 0, 0)                                                                                                                                                                \\
\multicolumn{1}{l}{}                                                                                                                                                   & \multicolumn{1}{l}{}                                                                                                                                                       \\
\end{longtable}

\begin{table*}
\caption{The diagonal elements of the stress tensor $\mathbf{\sigma}$, high frequency dielectric tensor $\mathbf{\epsilon^{\infty}}$ and averaged born effective charges $\mathbf{\bar{Z}}^*$  of the unique elements for the primitive cubic perovskite following the convention of reference \cite{Zhong1994}. For PZT, these calculations were performed at $a_{\text{Vg}}$ whilst PTO \& PZO were peromed at $a_{\text{PTO}}$ \& $a_{\text{PZO}}$ respectively.\\}  \medskip
\label{tab:BECDielecSigma}
\begin{tabular*}{500pt}{@{\extracolsep{\fill}}ccccccccccc}
              & $\sigma_{11}$ [GPa] & $\sigma_{22}$ [GPa] & $\sigma_{33}$ [GPa] & $\bar{Z}^*_{\text{Pb}}$ & $\bar{Z}^*_{\text{B}}$ & $\bar{Z}^*_{\text{O}_{\parallel}}$ & $\bar{Z}^*_{\text{O}_{\perp}}$ & $\epsilon^{\infty}_{11}$ & $\epsilon^{\infty}_{22}$ & $\epsilon^{\infty}_{33}$ \\ \toprule \toprule
\textbf{PTO}  & $\approx 0$         & $\approx 0$         & $\approx 0$         & 3.88                    & 7.19                   & -5.91                  & -2.58                          & 8.49                     & 8.49                     & 8.49                     \\
\textbf{PZO}  & $\approx 0$         & $\approx 0$         & $\approx 0$         & 3.90                    & 5.94                   & -4.90                  & -2.47                          & 6.93                     & 6.93                     & 6.93                     \\
\textbf{Mean} & $\approx 0$         & $\approx 0$         & $\approx 0$         & 3.89                    & 6.55                   & -5.41                  & -2.53                          & 7.71                     & 7.71                     & 7.71                     \\ \midrule
\textbf{VCA}  & -1.51               & -1.51               & -1.51               & 3.90                    & 6.13                   & -4.99                  & -2.52                          & 7.06                     & 7.06                     & 7.06                     \\
\textbf{I}    & 0.56                & 0.56                & 0.56                & 3.89                    & 6.54                   & -5.36                  & -2.53                          & 7.59                     & 7.59                     & 7.59                     \\
\textbf{II}   & 0.64                & -1.60               & 0.64                & 3.89                    & 6.55                   & -5.38                  & -2.53                          & 7.59                     & 7.67                     & 7.59                     \\
\bottomrule \bottomrule
\end{tabular*}
\end{table*}

    \subsection{\label{phonon:level2} Phonon dispersion and density of states}

\begin{figure*}
       \includegraphics[width=\linewidth]{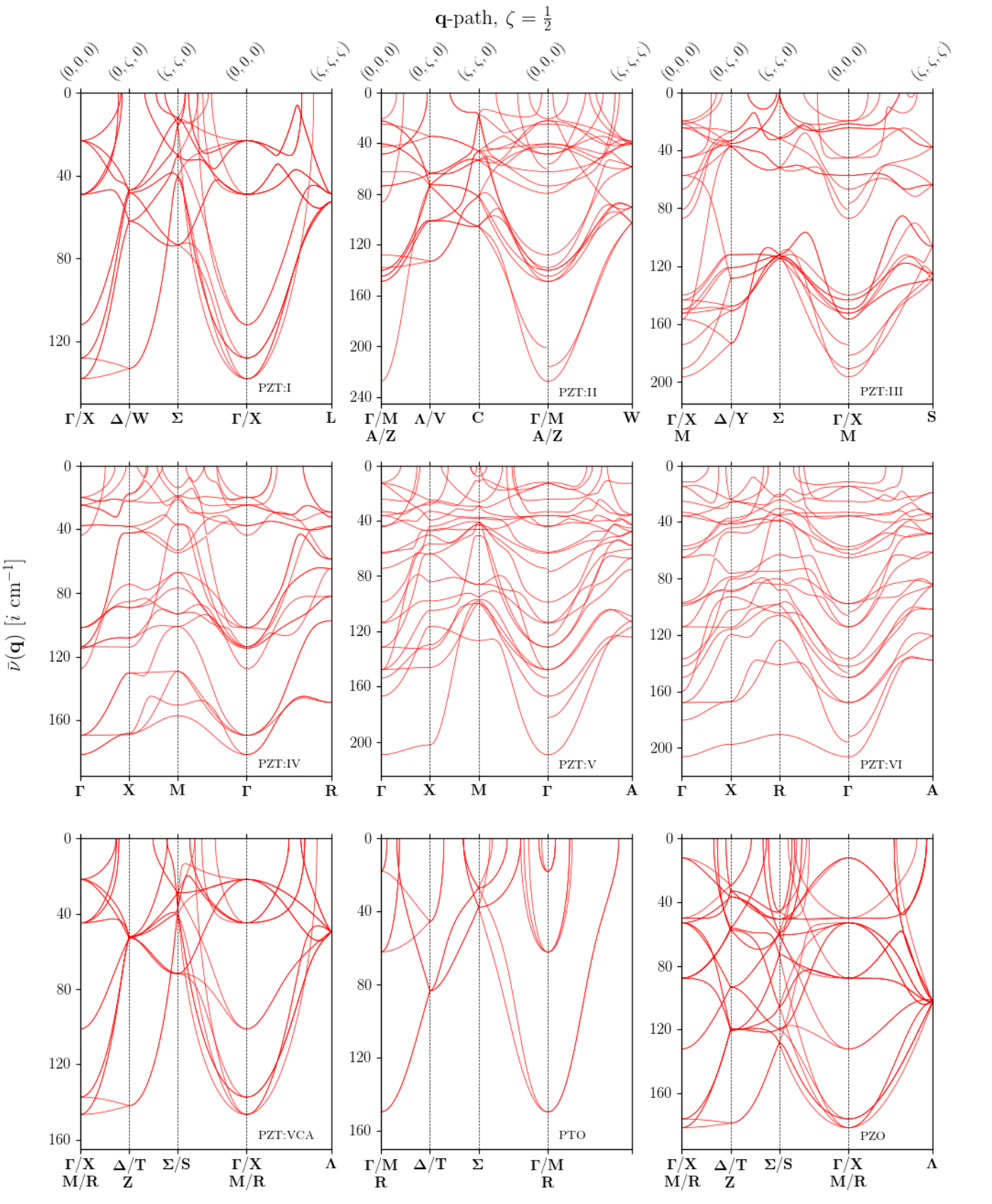}
        \caption{The soft mode phonon dispersion relations for PZT I:VI, the VCA, PTO and PZO. All dispersions are over an identical fractional $\mathbf{q}$-path controlled by the parameter $\zeta=1/2$ (upper $x$-axis). Folded symmetry labels are included for PZT I:III, the VCA, PTO and PZO. We only include folded labels if a soft mode of that wavevector is present at the given $\mathbf{q}$-point. Since dispersions for PZT IV:VI were calculated on the primitive cell, no folding takes place thus only one symmetry label is required.}
        
        \label{fig:dispersion}
    \end{figure*}

    Figure \ref{fig:dispersion} shows the phonon dispersions for PZT I:VI, PTO, PZO and the VCA calculated within the supercells indicated in figure \ref{fig:ConfigsPZT}. Although we have calculated all bands (available in section 3 of the Supplemental Material\cite{note:SM}), we consider only the space where $\bar{\nu}(\mathbf{q}) \in i\mathbb{R}$ thus presenting a set of symmetry lowering phase transitions along the fractional $\mathbf{q}$-path (0, 0, 0) $\Rightarrow$ (0, 1/2, 0) $\Rightarrow$ (1/2, 1/2, 0) $\Rightarrow$ (0, 0, 0) $\Rightarrow$ (1/2, 1/2, 1/2). It is at these points exactly that we analyze the character of the distortions. The soft mode character has an important impact on the properties of the of the crystal. This is then inferred with PDOS calculations (figures \ref{fig:FirstDOS} \& \ref{fig:SecondDOS}) and, for some important modes, found directly with eigendisplacement analysis.  Table \ref{tab:modes} serves as a companion to the dispersion identifying modes symmetries, their multiplicities and numerical values of imaginary frequencies. \par
    
    \subsubsection{\label{endmemeber:level3} PTO \& PZO}
    
    We begin with a discussion of end members PTO \& PZO. Our choice of supercell for these calculations reveals folded spectra not previously reported in the literature. We have also, however, calculated dispersions over the primitive cell and found good agreement with previous calculations using similar method \cite{Ghosez1999, Zhang2017} (see section 2 of the Supplemental Material\cite{note:SM}). For PTO we report 7 unique soft modes at the appropriate wavevectors compared to 26 in the more complex spectrum of PZO. As expected, the most unstable mode in PTO is found to be $\Gamma_4^-$ featuring Pb/Ti countermotion against the O anions inducing a net polarization and incipient FE distortion. Although the $\Gamma_4^-$ distortion exists in PZO, it is harder and features Zr motion \textit{alongside} O requiring that the smaller macroscopic polarization is as the result of Pb-O separation. PTO shows oxygen octahedron rotational instabilities at the R \& M points. These are the R$_4^+$ and M$_3^+$ AFD modes respectively. In real space, these correspond to out-of-phase and in-phase rotations of the BO$_6$ octahedra about a single axis, or a$^0$a$^0$c$^-$ \& a$^0$a$^0$c$^+$ in Glazer's notation, respectively. These modes are generally not competitive in PTO but this is \textit{not} true for PZO. The R$_4^+$ distortion is the softest mode in PZO and is a prime mover for the AFE phase transition known to make up $\approx$ 60\% of the total distortion \cite{inguez2014} (when the rotation is about the [$1\bar{1}0$] axis).\par 
    
    Branches mostly harden along the (0, 0, 0) $\Rightarrow$ (0, 1/2, 0) path in PTO resulting in an antipolar mode $\Delta_5$ and a long wavelength AFD mode T$_4$. The latter shares a likeness with both a$^0$a$^0$c$^-$ and a$^0$a$^0$c$^+$ distortions but with a doubled periodicity of four perovskite units along the axis of rotation. Of the four TiO$_6$ octahedra in the mode, two neighbouring octahedra rotate counterclockwise and the other two clockwise about the axis of rotation as seen in figure \ref{fig:FrozenImages}i (left). Although there is also a general hardening of branches along the same path in PZO, the softest is almost dispersionless resulting in another AFD mode of symmetry T$_4$. Although over the same wavevector as the T$_4$ mode of PTO, this mode is better described as a a$^0$a$^0$c$^-$-like distortion where rotating octahedra are separated by static ones (figure \ref{fig:FrozenImages}i, right). Both PTO and PZO now become harder at (1/2, 1/2, 0) resulting in several antipolar modes and for the first time in this study, single modes with a mixed antipolar/AFD character. These modes often manifest in a sublattice of BO$_6$ octahedra rotating with a Glazer-like pattern with adjacent PbBO$_3$ units showing local polar distortions. These local polar distortions are aligned such that there is no net polarization induced by the mode. An example of this is the $\Sigma_2$ distortion of PTO, although it has relatively low soft mode frequency (27.05$i$ cm$^{-1}$). Modes of this character are considerably softer in PZO including the S$_4$ distortion which features local AFD modes (with a complex non-Glazer-like rotation pattern) and antipolar cation displacements. This mode is also known to make a small contribution to the AFE PZO groundstate \cite{inguez2014}. \par
    
    Along the (0, 0, 0) $\Rightarrow$ (1/2, 1/2, 1/2) path, the dispersion now becomes real in PTO thus we see no instabilities at this longer wavelength. For PZO, the dispersion remains imaginary. We see a hardening resulting in two strongly degenerate modes of symmetry $\Lambda_2$ \& $\Lambda_3$. The former is an 8-fold degenrate AFD mode whilst the latter is 16-fold degenerate featuring Pb-O antipolar displacements. The character of these modes are reminscient of some of the known modes contributing to the PZO groundstate. This suggests that the inclusion of these distortions, with others, could create another similar low energy competing phase. Figure \ref{fig:FirstDOS} shows that the two end-members have a striking dissimilarity in the PDOS. All species for PTO show a rather featureless smooth function, peaking at $\approx$ 24$i$ cm$^{-1}$ whilst PZO shows a peaked PDOS penetrating further into the imaginary space indicating that cubic PZO is more dynamically unstable than PTO. The peak at $\approx$ 50$i$ cm$^{-1}$ is in part due to the dispersionless behaviour of a Pb-O antipolar branch extending from (1/2, 1/2, 0) $\Rightarrow$ (0, 0, 0). This behaviour continues for most of the (0, 0, 0) $\Rightarrow$ (1/2, 1/2, 1/2) path also. It is noteworthy that the Pb character vanishes for the softest part of the PZO PDOS leaving just modes of Zr-O character. \par
    
     \subsubsection{\label{VCAandI:level3} Virtual crystal approximation \& PZT I}
     
\begin{figure}
       \includegraphics[width=\linewidth]{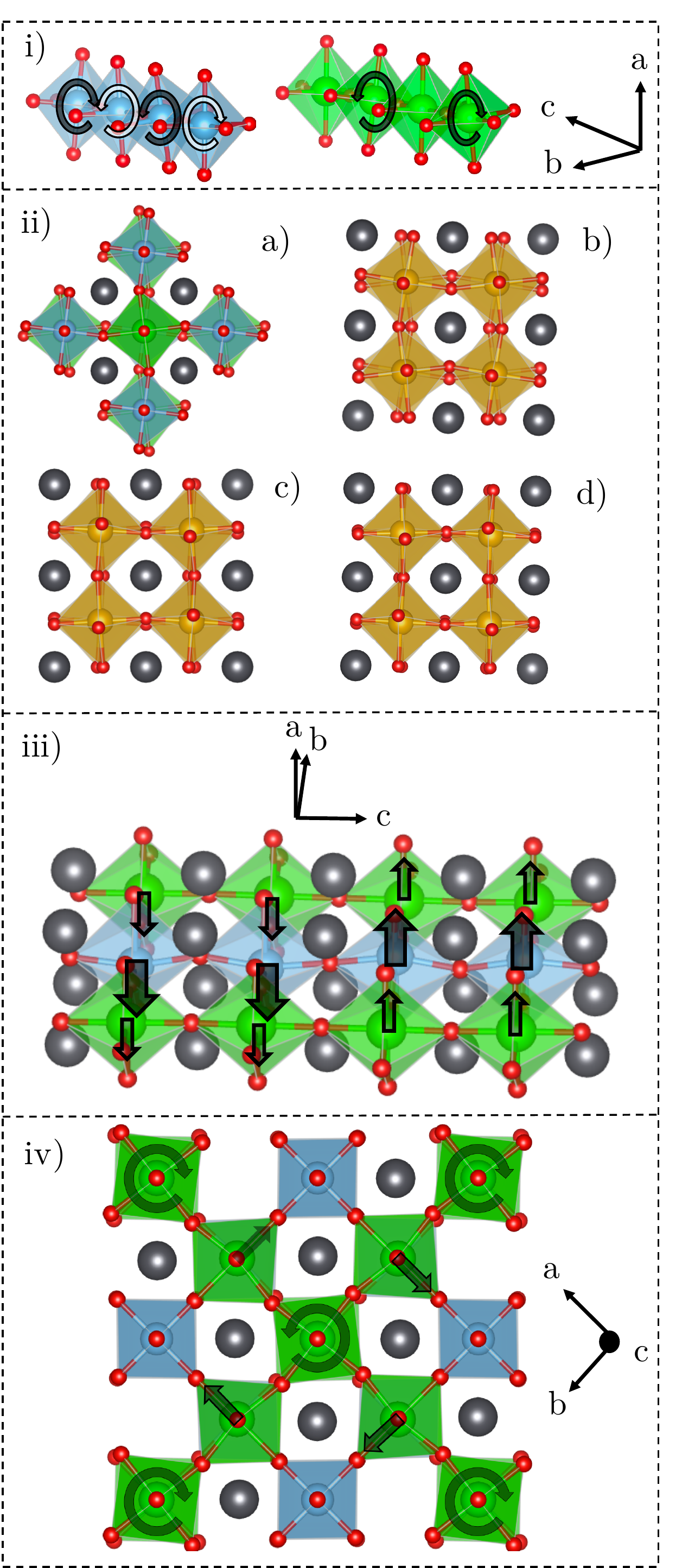}
        \caption{Visualisation of eigendisplacements described in the text following the same key as figure \ref{fig:ConfigsPZT} but also with grey spheres representing Pb sites. i) The T$_4$ modes of PTO and PZO. both Pb and counter-rotating octahedra are removed for clarity. ii) The $\Gamma_4^+$ distortion of PZT I (c-axes into page) and the M$_2^+$ distortion of the VCA from three viewing angles indicating out-of-phase rotation about three axes of rotation iii) The antipolar $\Delta_4$ distortion of PZT III. Arrows indicate the direction and magnitude of the local polarization. iv) the mixed antipolar/AFD M$_3^+$ distortion of PZT IV.}
        \label{fig:FrozenImages}
    \end{figure}

    There is a remarkable visual similarity in the dispersion relations between PZT I and the VCA. At first glance, this suggests that the within mixed potential scheme the dynamics of alternating Zr and Ti atoms in the rock-salt structure are well approximated. We do, however, see more unique branches for PZT I and find that the lowest lying modes of the VCA penetrate further into the soft space than its rock-salt ordered counterpart. It is also true that both approaches resemble PZO more so than PTO. This can be seen when assessing the modes at wavevector (0, 0, 0). At this point, PZT I, the VCA and PZO share a similar hierarchy of modes. PZT I and the VCA also share identical multiplicities. In descending order in imaginary wavenumber, we have out-of-phase AFD, in-phase AFD, FE then a number of antipolar modes. It is illuminating in this case, to perform a full analysis of the character. It soon becomes apparent that the VCA features AFD modes about all three axes of rotation. The amplitude of these rotations about two of the axes is small and much larger for the remaining axis. We could then consider these modes as rotations about a single axis but with small, erroneous rotations about the other axes. This is in contrast to PZT I where the softest AFD mode ($\Gamma_4^+$) has (like both end members) an a$^0$a$^0$c$^-$ displacement pattern, shown in figure \ref{fig:FrozenImages} ii) a). In the VCA, this rotation (M$_2^+$) retains its out-of-phase characteristic but now rotates about all three axes of rotation with different amplitudes thus exhibiting the a$^-$b$^-$c$^-$ rotation pattern shown in figure \ref{fig:FrozenImages} ii) b-d). The next softest mode in PZT I (X$_3^+$) has the  a$^0$a$^0$c$^+$ pattern whilst the in-phase rotations in the VCA (R$_5^-$), as before, have differing amplitudes about all three axes of rotations. This is the a$^+$b$^+$c$^+$ rotation pattern. The rotation patterns in the VCA are \textit{not} seen in any of the PZT supercell models indicating that rotations about more than one axis are a fictitious artifact of the method better illustrating the inaccuracy of the VCA in the prediction of local atomic displacements.  \par
    
\begin{figure}
       \includegraphics[width=\linewidth]{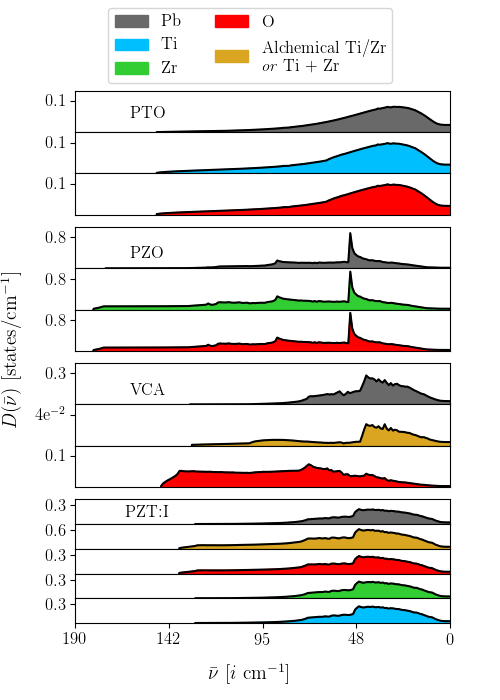}
        \caption{Species projected phonon density of states $D(\bar{\nu})$ for PTO, PZO the VCA and PZT I over the imaginary wavenumber space. For the VCA calculation, the gold curve is the PDOS of the alchemical 50/50 Ti/Zr atom whilst for the supercell models it represents the sum of B-site PDOS.}
        \label{fig:FirstDOS}
    \end{figure}

    The character of the FE $\Gamma_4^-$ modes are also dissimilar in nature. For PZT I, all ions play a role in the development of polarization including Ti and Zr displacement of a similar magnitude. For the VCA, the alchemical B-site plays much less of a role. We can then infer (without the full Berry-phase calculation) that the incipient polarization is smaller in magnitude in part owed to the smaller B-site displacements but also due to the smaller value of $\bar{Z}^*_B$ (table \ref{tab:BECDielecSigma}). A lack of alchemical B-site character is in fact common place for the VCA as evidenced in figure \ref{fig:FirstDOS} where although optically coupled to Pb motion, has an almost vanishingly small PDOS. This suggests that within the VCA, the B-site is dynamically inert. This leaves the softest modes of the VCA to have nearly a pure O character. It is only for Pb that we see similarity in the PDOS between the VCA and PZT I. We see a peak in the Pb $D(\bar{\nu})$ for PZT I at $\approx 48i$ cm$^{-1}$ which is shifted $\approx -2i$ cm$^{-1}$ in the VCA. The Pb peak in PZT I coincides with the other species. Such a coupling is in fact true for all PZT configurations and end members PTO/PZO. It is unique to the VCA that we see little coupling beween Pb \& O. This can be regarded as a knock-on effect of the inert B-site. Since Pb \& B-site vibrations are weakly coupled, the usual B-site displacements which would otherwise follow Pb are not present. It is these displacements which more greatly influence O motion since Pb has only a weaker mixed ionic/covalent interaction with O.

    \begin{table*}[]
    \caption{The 10 softest modes, for each structure, measured along the phonon dispersion path in FIG \ref{fig:dispersion} (with the exception of PTO, featuring only 7 soft modes over the dispersion path). Modes are listed in descending imaginary wavenumber $\bar{\nu}$ across the page. Each entry features a symmetry label for the irrep and a multiplicity M. Since PZT II, III, IV and VI feature directional polar modes, affected wavenumbers are given in the format $\bar{\nu}_{[010]}$/$\bar{\nu}_{[110]}$/$\bar{\nu}_{[111]}$. The full tabulation of all soft modes can be found in section 1 of the Supplemental Material\cite{note:SM}. \newline}
    \label{tab:modes}
    \begin{tabular*}{500pt}{@{\extracolsep{\fill}}cccccccccccc}
                         &               & M & $\bar{\nu}$ [$i$ cm$^{-1}$] &  &               & M & $\bar{\nu}$ [$i$ cm$^{-1}$] &  &               & M  & $\bar{\nu}$ [$i$ cm$^{-1}$] \\ \toprule \toprule
    \textbf{PTO}         & $\Gamma_4^-$  & 2 & 149.60                      &  & $\Delta_5^+$  & 4 & 83.40                       &  & R$_4^+$       & 3  & 62.12                       \\
                         & T$_4$         & 2 & 45.40                       &  & $\Sigma_3$    & 4 & 37.61                       &  & $\Sigma_2$    & 4  & 27.05                       \\
                         & M$_3^+$       & 3 & 18.02                       &  &               &   &                             &  &               &    &                             \\ \midrule
    \textbf{PZO}         & R$_4^+$       & 3 & 181.52                      &  & T$_4$         & 2 & 178.75                      &  & M$_3^+$       & 3  & 176.09                      \\
                         & $\Gamma_4^-$  & 2 & 132.14                      &  & S$_4$         & 4 & 128.32                      &  & T$_5$         & 4  & 120.45                      \\
                         & $\Sigma_2$    & 4 & 119.69                      &  & Z$_4$         & 4 & 119.53                      &  & S$_3$         & 4  & 105.37                      \\
                         & $\Lambda_2$   & 8 & 103.13                      &  &               &   &                             &  &               &    &                             \\ \midrule
    \textbf{VCA}         & M$_2^+$       & 3 & 146.58                      &  & $\Delta_5$    & 2 & 141.95                      &  & R$_5^-$       & 3  & 137.38                      \\
                         & $\Gamma_4^-$  & 2 & 101.18                      &  & $S_1$         & 4 & 71.78                       &  & T$_2$         & 4  & 52.87                       \\
                         & Z$_1$         & 4 & 52.18                       &  & T$_5$         & 4 & 52.14                       &  & $\Lambda_3$   & 16 & 49.55                       \\
                         & X$_5^-$       & 6 & 44.89                       &  &               &   &                             &  &               &    &                             \\ \midrule
    \textbf{I}           & $\Gamma_4^+$  & 3 & 138.10                      &  & $\Delta_4$    & 2 & 133.08                      &  & X$_3^+$       & 3  & 128.11                      \\
    \multicolumn{1}{l}{} & $\Gamma_4^-$  & 2 & 111.99                      &  & $\Sigma_2$    & 4 & 73.53                       &  & $\Delta_5$    & 4  & 62.02                       \\
    \multicolumn{1}{l}{} & L$_3^-$       & 8 & 52.83                       &  & X$_5^-$       & 6 & 49.09                       &  & L$_3^{-\prime}$      & 8  & 48.70                       \\
                         & W$_5$         & 4 & 48.29                       &  &               &   &                             &  &               &    &                             \\ \midrule
    \textbf{II}          & M$_3^-$       & 1 & 227.52                      &  & $\Gamma_5^-$  & 1 & 144.78/201.13/215.93        &  & Z$_5^-$       & 2  & 148.78                      \\
                         & A$_5^-$       & 2 & 140.02                      &  & Z$_1^-$       & 1 & 137.91                      &  & $\Lambda_4$   & 2  & 132.92                      \\
                         & $\Gamma_3^+$  & 1 & 127.97                      &  & C$_1$         & 4 & 105.26                      &  & $\Lambda_5$   & 4  & 101.14                      \\
                         & W$_2$         & 4 & 90.22                       &  &               &   &                             &  &               &    &                             \\ \midrule
    \textbf{III}         & $\Gamma_5^-$  & 1 & 196.63/196.63/196.63        &  & M$_2^+$       & 1 & 190.76                      &  & $\Gamma_3^-$  & 1  & 174.29/174.29/181.62        \\
                         & $\Delta_4$    & 2 & 173.52                      &  & X$_2^+$       & 2 & 156.67                      &  & M$_5^+$       & 2  & 152.61                      \\
                         & $\Delta_3$    & 2 & 150.97                      &  & $X_3^+$       & 2 & 149.79                      &  & Y$_3$         & 2  & 147.85                      \\
                         & X$_2^-$       & 2 & 143.46                      &  &               &   &                             &  &               &    &                             \\ \midrule
    \textbf{IV}          & $\Gamma_4^-$  & 2 & 181.41                      &  & $\Gamma_4^+$  & 3 & 169.33                      &  & X$_3^+$       & 1  & 168.90                      \\
                         & X$_5^+$       & 2 & 168.18                      &  & M$_2^-$       & 1 & 157.00                      &  & M$_3^+$       & 1  & 150.48                      \\
                         & R$_4^+$       & 3 & 148.69                      &  & X$_5^{+\prime}$      & 2 & 130.13                      &  & M$_5^+$       & 2  & 129.20                      \\
                         & $\Gamma_4^{-\prime}$ & 2 & 114.61                      &  &               &   &                             &  &               &    &                             \\ \midrule
    \textbf{V}           & $\Gamma_5^-$  & 1 & 209.42/209.42/209.42        &  & X$_3^-$       & 1 & 202.33                      &  & $\Gamma_3^-$  & 1  & 153.93/153.93/182.48        \\
                         & $\Gamma_3^+$  & 1 & 167.07                      &  & $\Gamma_5^{-\prime}$ & 1 & 148.18/148.18/123.20        &  & $\Gamma_5^+$  & 1  & 147.67                      \\
                         & X$_4^+$       & 1 & 146.16                      &  & X$_1^-$       & 1 & 132.83                      &  & $\Gamma_5^{+\prime}$ & 2  & 131.48                      \\
                         & X$_2^-$       & 1 & 129.84                      &  &               &   &                             &  &               &    &                             \\ \midrule
    \textbf{VI}          & $\Gamma_5^-$  & 1 & 206.22/206.22/206.22        &  & X$_2^+$       & 1 & 197.36                      &  & $\Gamma_3^-$  & 1  & 180.33/195.87/191.69        \\
                         & R$_1^-$       & 1 & 190.45                      &  & $\Gamma_5^+$  & 2 & 167.85                      &  & X$_4^+$       & 1  & 167.42                      \\
                         & X$_3^+$       & 1 & 167.00                      &  & $\Gamma_5^{-\prime}$ & 1 & 159.78/146.71/150.21        &  & $\Gamma_4^+$  & 1  & 149.99                      \\
                         & $\Gamma_3^+$  & 1 & 142.07                      &  &               &   &                             &  &               &    &    \\  \bottomrule \bottomrule                        
    \end{tabular*}
    \end{table*}

    Moving away from (0, 0, 0) towards (0, 1/2, 0) both PZT I and the VCA give rise to longer wavelength AFD and antipolar modes. The T$_2$ and $\Delta_4$ modes of the VCA and PZT I, respectively, display the same rotation pattern as the aformentioned T$_4$ distortion in PTO. This mode is significantly more unstable in PZT I. Despite the VCA appearing to have a higher degeneracy for the antipolar soft modes at $\approx 50i$ cm$^{-1}$, modes are still unique splitting only by $\approx 0.5i$ cm$^{-1}$. One of these modes, Z$_1$, is not purely antipolar and once again we see the mixed AFD/antipolar character displaying non-Glazer-like rotations coupled with Pb cation motion.\par
    
    Like PZO, both the VCA and PZT I become their hardest along the (0, 1/2, 0) $\Rightarrow$ (1/2, 1/2, 0) path. This leads to further antipolar modes at the $\Sigma$ and S points. Notably, PZT I gains an additional soft mode from the real domain along this path, $\Sigma_1$. This is distinct from the other Pb-O modes since it features antipolar Pb-B displacements with no significant O character. For the VCA, there is also a 4-fold degenerate mode Z$_1$ once again with mixed antipolar/AFD character. The most distinct differences in the dynamical behaviour between the VCA and PZT I now comes along the path (0, 0, 0) $\Rightarrow$ (1/2, 1/2, 1/2). Many of the harder antipolar branches in the VCA move to the real domain. These modes do begin to harden in PZT I but then re-soften to become degenerate with other branches at the L-point giving rise to two long wavelength modes both of symmetry L$_3^-$. Now commonplace, they share a mixed antipolar/AFD character split by $\Delta \bar{\nu} = 4.13i$ cm$^{-1}$. We distinguish between modes sharing an irrep by priming those with the lower imaginary frequency as seen in table \ref{tab:modes}. Each mode has 8-fold degeneracy despite L$_3^{-\prime}$ having a longer wavelength AFD rotation pattern than L$_3^-$. This splitting closes for the VCA giving rise to one 16-fold degenerate mode of symmetry $\Lambda_3$ displaying a similar mixed antipolar/AFD character. \par 
    
     \subsubsection{\label{IIandIII:level3} PZT II \& III}
     
     We move now to consider the dispersions of PZT II \& III. These are the [110] and [001] ordered superlattices respectively. These structures were considered in a previous work in a study of the instabilities at the $\Gamma$-point \cite{Bungaro2002} using the local density approximation (LDA). Consistent with the previous work, we find that both PZT II \& III have strong TO FE instabilities of $\Gamma_3^-$ \& $\Gamma_5^-$ symmetry respectively. The softest TO mode of the [110] ordered structure is not seen in our dispersion path due to the anisotropy of LO-TO splitting in non-cubic crystals. This anisotropy can be reasoned by the form of the NAC. Recall that the NAC is a function of both $\mathbf{Z}^*_i$ \& $\mathbf{\epsilon}^{\infty}$. The former gains more unique elements in lower symmetry crystals and the latter is no longer isotropic as evidenced in table \ref{tab:BECDielecSigma}. The affected elements of the dynamical matrix are then corrected by a different amount based on the direction of the $\mathbf{q}$-vector as it approaches $\Gamma$. This effect is seen in PZT II, II, V \& VI since they are all members of a tetragonal spacegroup. Taking just the analytic part of the $\Gamma_3^-$ mode of PZT II returns an eigenfrequency of 242.28$i$ cm $^{-1}$, slightly softer than what is predicted by the LDA. \par 
     
     We find that both PZT II \& III give rise to soft LO modes, again, in agreement with the previous work. [110] ordering is generally more dynamically unstable than [001] ordering showing a distinct separation between the most imaginary FE/antipolar modes and groupings of Glazer AFD modes. What was not considered in a previous study \cite{Bungaro2002} was competition of polar modes with other order parameters. The antipolar mode M$_3^-$ of PZT II is closely competitive with $\Gamma_3^-$. This mode is an antipolar arrangement of Ti-O displacements completely isolated to local PTO environments, leaving undistorted areas of PZO units. There are also a plethora of unique Glazer tilt modes owed to inequivalent directions in the crystals and thus inequivalent axes of rotation. The softest of these is an a$^0$a$^0$c$^-$ mode with the axes of rotation along the [001] (or [010]) direction, the direction of compositional modulation. This is followed by a several antipolar modes and harder FE modes. In PZT III, rotational instability is highly competitive with FE order due to the M$_2^+$ mode. This mode shows in-phase rotation of ZrO$_6$ octahedra, leaving the TiO$_6$ octahedra static in a manner reminiscent of the T$_4$ distortion of PZO. This shows there is no mechanical coupling along the axis of rotation between octahedra centered on a different B-site species. Whilst rotations of all octahedra are also unstable (both out-of-phase M$_5^+$ and in-phase X$_2^-$), they are harder. Further, both of these modes rotate along homogeneous B-site chains whereas the M$_2^+$ mode rotates along the heterogeneous direction where no other Glazer type instability exists. \par 
     
    The character of AFD modes in PZT II alters as we approach the wavevector (0, 1/2, 0). This mode shows out-of-phase rotations of the ZrO$_6$ octahedra but with a doubled periodicity. Rotating octahedra are also separated by static ZrO$_6$ octahedra this time showing a lack of inter-layer coupling even along the homogeneous direction. A long wavelength AFD mode also exists for PZT III at this wavevector of irrep Y$_3$. This modes shows the same character of the T$_4$ mode of PTO with the axis of rotation being along the homogeneous direction. This wavevector for PZT III, however, is dominated by antipolar instability with the most unstable being the $\Delta_4$ mode. This mode appears with two separate polar domains with a domain period of 4 perovskite units, separated by a 180$^{\circ}$ domain wall as depicted in figure \ref{fig:FrozenImages} iii). Local PTO units are significantly more polar than local PZO units. /par
    
    Like in PZT I, the VCA, PTO and PZO, the most imaginary bands at (0, 1/2, 0) have a steep gradient to the hard wavevector (1/2, 1/2, 0). This results in tight groupings of antipolar and mixed antipolar/AFD modes for PZT III but only antipolar modes for PZT II. The dispersion now returns to (0, 0, 0). We note that along this direction of approach ([110]), anisotropy in LO-TO splitting allows for softer LO FE modes to appear in both PZT II \& III and softer still along the [111] direction. This results in a sharp discontinuities in the spectra. From (0, 0, 0) to the long wavelength (1/2, 1/2, 1/2) point, hardening occurs for both PZT II \& III giving rise to 5 distinct distortions for each arrangement. For PZT II, these are the W$_{1:4}$ (where the subscript indicates all modes with integers 1 through 4) and W$_1^{\prime}$ distortions. Each of these modes has a pure antipolar character. further, the splitting of the isosymmetrical modes W$_1$ and W$_1^{\prime}$ is large (64.16$i$ cm$^{-1}$) due to the inclusion of Zr displacement in  W$_1^{\prime}$ where W$_1$ features static Zr. PZT III possesses similar characteristics in its long wavelength distortions, S$_{1:4}$ and S$_4^{\prime}$. Unlike PZT II, two of these distortions have the mixed AFD/antipolar character whilst the remaining are purely antipolar. The S$_4$-S$_4^{\prime}$ splitting is also large (61.19$i$ cm$^{-1}$) but is now the result of the inclusion of local AFD displacements in S$_4$ whilst S$_4\prime$ is purely antipolar. \par
    
\begin{figure}
       \includegraphics[width=\linewidth]{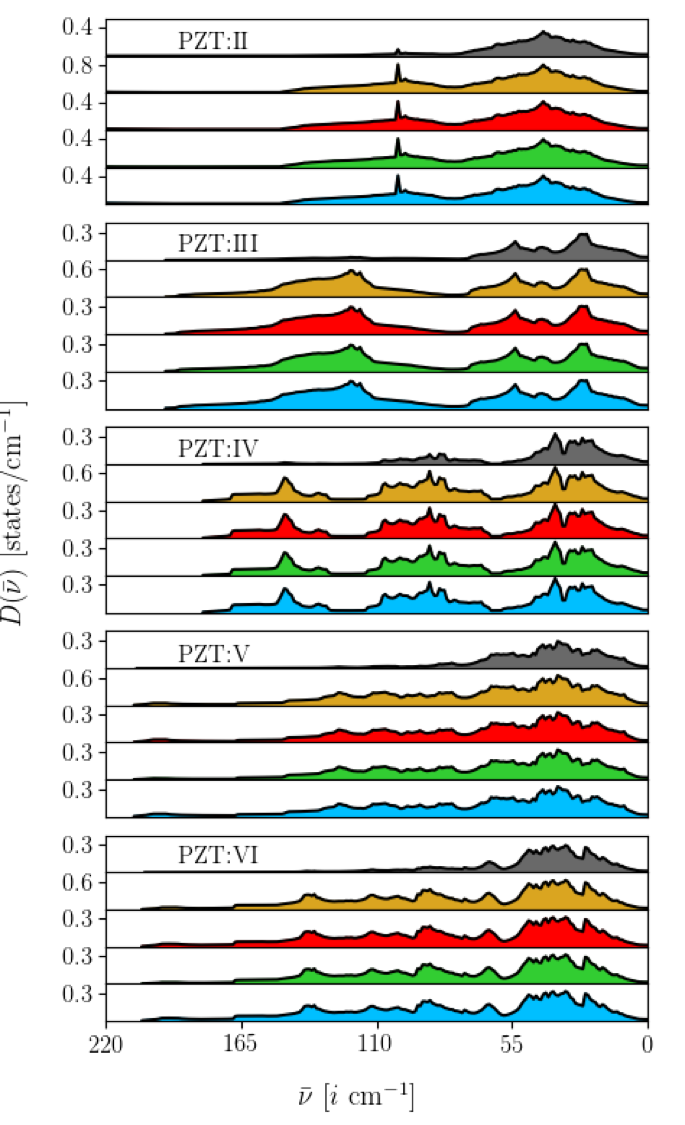}
        \caption{Species projected phonon density of states $D(\bar{\nu})$ for PZT III:VI over the imaginary wavenumber space. This figure shares a legend with FIG \ref{fig:FirstDOS}.}
        \label{fig:SecondDOS}
    \end{figure}

    The general character of the disortions in both PZT II \& III can be inferred from the PDOS (figure \ref{fig:SecondDOS}). We see that for both arrangements, all species are optically coupled to one-another, but, like before, the Pb character starts to diminish as we penetrate further into the soft domain. Whilst both PZT II \& III both give rise to two separated islands of states in the PDOS, a sharp peak exists on the softer island of PZT II at $\approx$ 100$i$ cm$^{-1}$. This is owing to the nearly dispersionless behaviour of the antipolar branch connecting the $\Lambda_5$ and C$_1$ modes. The 4-fold degenerate W$_1$ anti-polar mode also appears at this wavenumber (along with $\Lambda_5$ and C$_1$) containing significant Pb character.\par

      \subsubsection{\label{IV:level3} PZT IV}
     
     We discuss now PZT IV in isolation, which, despite sharing $m\bar{3}m$ symmetry with PZT I, shows radically different dynamical behaviour as well as being generally more unstable. For the first time in this study, also, we consider dispersion over what is the primitive lattice so we pass through high symmetry points without any folding of the BZ. Unlike PZT I, the softest mode at (0, 0, 0) is now a $\Gamma_4^-$ distortion suggesting a FE groundstate. This distortion shows stronger local polarity in directions with continuous PTO units. The presence of Zr along a polar direction dampens the distortion. For the first time in this study, no pure Glazer type AFD instabilities are found to exist in a single mode. These are replaced with isolated in-phase AFD instabilities the softest of which is the $\Gamma_4^+$ mode. This mode features a rotating layer (isolated by static PbBO$_3$ layers) with a ratio of 8:1 ZrO$_6$ to TiO$_6$ octahedra. In this case, the dominance of the PZO rich environment (which favours rotation) is able to overpower the single PTO unit (favouring FE distortion) into rotation. One other rotational instability exists at this point, $\Gamma_4^{+ \prime}$. This mode shares the same characteristics as $\Gamma_4^+$, but the rotating layer contains fewer ZrO$_6$ octahedra making the mode more stable than its counterpart. It is notable that there are three separate occurrences of the FE $\Gamma_4^-$ irrep: $\Gamma_4^-$, $\Gamma_4^{-\prime}$ and $\Gamma_4^{-\prime \prime}$. The latter (although much harder than the others) is distinct not only due to its weak B-site displacements but alternating Pb cation motion transverse to the direction of polarization giving rise to a mode of a mixed FE \& antipolar character at the zone centre. \par 
     
     Most bands harden only slightly along the path to X much in contrast to the superlattice type arrangements. Antipolar type distortions at this wavevector are much harder than previous arrangements featuring only Pb-O motion. There is now only a slight hardening in the dispersion along the $\Gamma \Rightarrow$ X path once again leading to a selection of antipolar and AFD modes. The X$_3^+$ and X$_5^+$ modes are particularly unstable. The first is a long wavelength AFD mode much like $\Gamma_4^+$ but with out-of-phase rotations. These rotational modes are very closely competing split by $<$1$i$ cm$^{-1}$ in the favour of X$_3^+$. The second, X$_5^+$ is an isolated antipolar distortion where local PTO units are polar in the direction of compositional homogeneity. PZO units are once again resistant to polarization and are left static. After a small degree of hardening along the path to M, we find 15 unique distortions of antipolar and mixed AFD/antipolar character; the largest concentration oh such states in this study. The softest is antipolar M$_2^-$ bearing great resemblance to X$_5^+$ but over a greater wavelength.
     
     The M$_3^+$ mode is the clearest example of a mixed AFD/antipolar mode. This is shown in figure \ref{fig:FrozenImages} iv). It features a central in-phase rotation similar to $\Gamma_4^+$. PbBO$_3$ units perpendicular to the axis of rotation now show local polar displacements in a pattern enclosing the central rotating unit. Softer modes of this character can be seen at the R-point. Here we find that the most unstable branches are dominated by the mixed AFD/antipolar character. In-fact, the unstable mode of this character, amongst all PZT arrangements, is found here and is the triply degenerate R$_4^+$. This shares great similarity to M$_3^+$ but rotations are out of phase and about two axes making the rotation pattern a$^0$b$^-$b$^-$-like. Other modes at this wavevector are also visually similar to M$_3^+$ but now the local polar regions include Pb \& Ti cation motion where before local polarity was just as the result of O displacing against static Zr. \par
     
     PZT IV is the only arrangement to form three distinct islands in the PDOS. The two more stable islands feature coupled ionic motion between all species, but, as before the most imaginary states have a diminished Pb character. It is clear that the first (and least imaginary) island is comprised entirely of antipolar states and the second of antipolar and mixed AFD/antipolar states. The softest island features the purely rotational states but also FE and mixed AFD/antipolar order. Unlike previous arrangements, the is a significant peak in the most unstable island at $\approx$ 140$i$ cm$^{-1}$ as a result of a significant amount of mixed AFD/antipolar modes. This suggests that such a mode character could play a role in a low energy structure of this arrangement.
     
     \subsubsection{\label{VandVI:level3} PZT V \& VI}
     
     The last of the arrangements we consider together are PZT V \& VI. A striking dissimilarity between these two arrangements and the rest is the increased number on unique bands in the soft space. The vast majority of these states are singly degenerate in response to the large number of uniquely coordinated ions. At the $\Gamma$-point, both arrangements are dominated by a highly imaginary FE distortion of symmetry $\Gamma_5^-$. Both distortions display greater local polarization in the direction of compositional homogeneity in Ti. Local PZO units are polarized but as in the end member PZO, Zr play less of a role. Both arrangements feature other polar modes where like PZT IV, Pb cation motion is in a direction perpendicular to the polarization soggesting a dual FE \& antipolar character. \par
     
     Like PZT III and IV, PZT V favours isolated rotations separated by static octahedra. One example of this is $\Gamma_3^+$ mode where rotating layers feature a higher number of Zr sites and static layers have a higher number of Ti sites. It is true once more that purely Glazer type rotations are not seen in the spectra of PZT V. These are replaced with Glazer-like modes where one layer rotates more strongly than the other. The most unstable example of this is the $\Gamma_5^+$ mode which is strongly a$^0$a$^0$c$^-$-like, but, the rotating layer with the higher Ti/Zr ratio rotates at a diminished amplitude. For both V \& VI, the most imaginary polar branch is almost dispersionless along the path to X resulting in the softest mode at the X$_3^-$ and X$_2^+$ for each arrangement, respectively. Both modes are antipolar featuring no Pb cation motion but heavy Ti-O countermotions. Like previous PZT arrangements at this wavevector, we see non-Glazer-like isolated AFD modes and a variety of harder antipolar modes. \par
     
     For PZT VI, we see that the most imaginary TO branch is not only dispersionless along the previously mentioned path but is for much of the BZ until we see a rapid hardening as we approach $\Lambda$. Even here however, the branch remains unstable. This shares some similarity with the dynamical behaviour seen in the dispersion relations of BaTiO$_3$ (BTO), but, for BTO the result is a confinement of the instability to three quasi-two-dimensional slabs of $\mathbf{q}$-space intersecting at $\Gamma$ since the branch becomes real towards the R-point. With the exception of this branch, the character of modes at the wavevectors (1/2, 1/2, 0) \& (1/2, 1/2, 1/2) are rather similar. Both give rise to large number of unique AFD/antipolar distortions similar to those described before. Notable also is the anisotropic behaviour of polar branches approaching the the $\Gamma$-point from the different considered directions. Whilst the most imaginary TO branches are unaffected, discontinuity can be seen clearly when comparing the [110] \& [111] directions for both PZT V \& VI which is tabulated in table \ref{tab:modes}. The farily even distribution of states across the soft space results in a single island in the PDOS for both PZT V \& VI, although, like other PZT arrangements, there is a higher density of antipolar states in the harder part of the soft-space. Remakably, despite the near-dispersionless character of the most imaginary polar branch in PZT VI, the resulting peak in the PDOS is small as a result of its isolation from other bands in the spectra and its single-fold degeneracy.  

     

\section{\label{summary:level1} Summary}
 We have explored the soft mode lattice dynamics of PTO, PZO and PbZr$_{0.5}$Ti$_{0.5}$O$_3$ and determined the character of the most unstable modes of each arrangement. This has revealed a complex landscape of local minima and possible phase transition paths for each arrangement. It is important to ephasize that this work indicates that altering B-site ordering in a fixed concentration of Ti/Zr in PZT can in some cases lead to the dominance of different order parameters. We find that, in general, (with the exception of PZT IV) that higher symmetry supercells and the VCA are dominated by rotational instabilities of the BO$_6$ octahedra which, like pure PZO, are able to couple with Pb antipolar modes at the $\Sigma$-point suggesting the stability of an AFE structure. Lower symmetry supercells are found to be more PTO-like implicated by the soft zone centre modes with a FE character. Given the number of unique distortions, however, it is unlikely that the ground state of these structures can be described only by a FE distortion. Further work could include identification of the phase transition paths resulting in a mixed-mode groundstate. \par
 
 We find also that there is considerable competition with the routinely considered polar and Glazer-like rotational modes from longer wavelength antipolar modes and with non-Glazer-like AFD modes. In some cases non-Glazer-like isolated out-of-phase rotation of ZrO$_6$ octahedra is more unstable than Glazer a$^{0}$a$^{0}$c$^+$ and is either closely competitive with or more unstable than a$^{0}$a$^{0}$c$^-$ distortions. For PZT IV \& V, we find \textit{no} soft modes which result Glazer type rotations. We find that some soft modes can give rise to distortions characteristic of more than one order parameter. It is found that in PZT IV:VI that FE order can appear simultaneously with antipolar Pb displacements. All PZT arrangements have long wavelength soft modes displaying a dual antipolar/AFD character. It is possible that such distortions are competitive in PZT IV suggesting a complex local minima rivalling the softer FE distortion. Given the long wavelengths associated with these modes, there are a large number of participating atoms. It can then become costly to study their behaviour with conventional plane-wave based DFT due to well known scaling issues. Accurate first principles simulations of these systems will then require large-scale electronic structure methods \cite{Bowler2012}. \par
 
 The applicability of the VCA as a substitute for the supercell method has been investigated. Whilst the disperion looks strikingly similar to that of PZT I, we find that the species specific character is considerably different. The alchemical Ti/Zr atom does not play a role in the lattice dynamics but rather is a site inert to displacement. Crucially, the softest Glazer type rotational modes have a different classication in the VCA becoming a$^{-}$b$^{-}$c$^{-}$ \& a$^{+}$b$^{+}$c$^{+}$ as opposed to a$^{0}$a$^{0}$c$^{-}$ \& a$^{0}$a$^{0}$c$^{+}$ like found in other PZT supercells and end members PTO \& PZO. This quantitatively displays the inability of the VCA to represent local structural distortions.

\section{\label{Acknowledgements:level1} Acknowledgements}
We are grateful for computational support from the UK Materials and Molecular Modelling Hub, which is partially funded by EPSRC (EP/P020194), for which access was obtained via the UKCP consortium and funded by EPSRC grant ref EP/P022561/1. This work also used the ARCHER UK National Supercomputing Service (http://www.archer.ac.uk) funded by the UKCP consortium EPSRC grant ref EP/P022561/1.

\bibliography{main}

\begin{thebibliography}{62}%
\makeatletter
\providecommand \@ifxundefined [1]{%
 \@ifx{#1\undefined}
}%
\providecommand \@ifnum [1]{%
 \ifnum #1\expandafter \@firstoftwo
 \else \expandafter \@secondoftwo
 \fi
}%
\providecommand \@ifx [1]{%
 \ifx #1\expandafter \@firstoftwo
 \else \expandafter \@secondoftwo
 \fi
}%
\providecommand \natexlab [1]{#1}%
\providecommand \enquote  [1]{``#1''}%
\providecommand \bibnamefont  [1]{#1}%
\providecommand \bibfnamefont [1]{#1}%
\providecommand \citenamefont [1]{#1}%
\providecommand \href@noop [0]{\@secondoftwo}%
\providecommand \href [0]{\begingroup \@sanitize@url \@href}%
\providecommand \@href[1]{\@@startlink{#1}\@@href}%
\providecommand \@@href[1]{\endgroup#1\@@endlink}%
\providecommand \@sanitize@url [0]{\catcode `\\12\catcode `\$12\catcode
  `\&12\catcode `\#12\catcode `\^12\catcode `\_12\catcode `\%12\relax}%
\providecommand \@@startlink[1]{}%
\providecommand \@@endlink[0]{}%
\providecommand \url  [0]{\begingroup\@sanitize@url \@url }%
\providecommand \@url [1]{\endgroup\@href {#1}{\urlprefix }}%
\providecommand \urlprefix  [0]{URL }%
\providecommand \Eprint [0]{\href }%
\providecommand \doibase [0]{http://dx.doi.org/}%
\providecommand \selectlanguage [0]{\@gobble}%
\providecommand \bibinfo  [0]{\@secondoftwo}%
\providecommand \bibfield  [0]{\@secondoftwo}%
\providecommand \translation [1]{[#1]}%
\providecommand \BibitemOpen [0]{}%
\providecommand \bibitemStop [0]{}%
\providecommand \bibitemNoStop [0]{.\EOS\space}%
\providecommand \EOS [0]{\spacefactor3000\relax}%
\providecommand \BibitemShut  [1]{\csname bibitem#1\endcsname}%
\let\auto@bib@innerbib\@empty
\bibitem [{\citenamefont {Jaffe}(2012)}]{jaffe2012piezoelectric}%
  \BibitemOpen
  \bibfield  {author} {\bibinfo {author} {\bibfnamefont {B.}~\bibnamefont
  {Jaffe}},\ }\href@noop {} {\emph {\bibinfo {title} {Piezoelectric
  ceramics}}},\ Vol.~\bibinfo {volume} {3}\ (\bibinfo  {publisher} {Elsevier},\
  \bibinfo {year} {2012})\BibitemShut {NoStop}%
\bibitem [{\citenamefont {Oliveira}\ \emph {et~al.}(2014)\citenamefont
  {Oliveira}, \citenamefont {Longo}, \citenamefont {Varela},\ and\
  \citenamefont {Zaghete}}]{Oliveira2014}%
  \BibitemOpen
  \bibfield  {author} {\bibinfo {author} {\bibfnamefont {C.}~\bibnamefont
  {Oliveira}}, \bibinfo {author} {\bibfnamefont {E.}~\bibnamefont {Longo}},
  \bibinfo {author} {\bibfnamefont {J.}~\bibnamefont {Varela}}, \ and\ \bibinfo
  {author} {\bibfnamefont {M.}~\bibnamefont {Zaghete}},\ }\href {\doibase
  10.1016/j.ceramint.2013.07.068} {\bibfield  {journal} {\bibinfo  {journal}
  {Ceramics International}\ }\textbf {\bibinfo {volume} {40}},\ \bibinfo
  {pages} {1717} (\bibinfo {year} {2014})}\BibitemShut {NoStop}%
\bibitem [{\citenamefont {Izyumskaya}\ \emph {et~al.}(2007)\citenamefont
  {Izyumskaya}, \citenamefont {Alivov}, \citenamefont {Cho}, \citenamefont
  {Morko{\c{c}}}, \citenamefont {Lee},\ and\ \citenamefont
  {Kang}}]{Izyumskaya2007}%
  \BibitemOpen
  \bibfield  {author} {\bibinfo {author} {\bibfnamefont {N.}~\bibnamefont
  {Izyumskaya}}, \bibinfo {author} {\bibfnamefont {Y.-I.}\ \bibnamefont
  {Alivov}}, \bibinfo {author} {\bibfnamefont {S.-J.}\ \bibnamefont {Cho}},
  \bibinfo {author} {\bibfnamefont {H.}~\bibnamefont {Morko{\c{c}}}}, \bibinfo
  {author} {\bibfnamefont {H.}~\bibnamefont {Lee}}, \ and\ \bibinfo {author}
  {\bibfnamefont {Y.-S.}\ \bibnamefont {Kang}},\ }\href {\doibase
  10.1080/10408430701707347} {\bibfield  {journal} {\bibinfo  {journal}
  {Critical Reviews in Solid State and Materials Sciences}\ }\textbf {\bibinfo
  {volume} {32}},\ \bibinfo {pages} {111} (\bibinfo {year} {2007})}\BibitemShut
  {NoStop}%
\bibitem [{\citenamefont {Gururaja}\ \emph {et~al.}(1985)\citenamefont
  {Gururaja}, \citenamefont {Schulze}, \citenamefont {Cross}, \citenamefont
  {Newnham}, \citenamefont {Auld}, \citenamefont {Wang} \emph
  {et~al.}}]{gururaja1985piezoelectric}%
  \BibitemOpen
  \bibfield  {author} {\bibinfo {author} {\bibfnamefont {T.}~\bibnamefont
  {Gururaja}}, \bibinfo {author} {\bibfnamefont {W.~A.}\ \bibnamefont
  {Schulze}}, \bibinfo {author} {\bibfnamefont {L.~E.}\ \bibnamefont {Cross}},
  \bibinfo {author} {\bibfnamefont {R.~E.}\ \bibnamefont {Newnham}}, \bibinfo
  {author} {\bibfnamefont {B.~A.}\ \bibnamefont {Auld}}, \bibinfo {author}
  {\bibfnamefont {Y.~J.}\ \bibnamefont {Wang}},  \emph {et~al.},\ }\href@noop
  {} {\bibfield  {journal} {\bibinfo  {journal} {IEEE Trans. Sonics Ultrason}\
  }\textbf {\bibinfo {volume} {32}},\ \bibinfo {pages} {481} (\bibinfo {year}
  {1985})}\BibitemShut {NoStop}%
\bibitem [{\citenamefont {Chi}\ and\ \citenamefont {Xu}(2014)}]{Chi2014}%
  \BibitemOpen
  \bibfield  {author} {\bibinfo {author} {\bibfnamefont {Z.}~\bibnamefont
  {Chi}}\ and\ \bibinfo {author} {\bibfnamefont {Q.}~\bibnamefont {Xu}},\
  }\href {\doibase 10.5772/59099} {\bibfield  {journal} {\bibinfo  {journal}
  {International Journal of Advanced Robotic Systems}\ }\textbf {\bibinfo
  {volume} {11}},\ \bibinfo {pages} {182} (\bibinfo {year} {2014})}\BibitemShut
  {NoStop}%
\bibitem [{\citenamefont {Kadin}\ and\ \citenamefont
  {Kaplan}(2017)}]{Kadin2017}%
  \BibitemOpen
  \bibfield  {author} {\bibinfo {author} {\bibfnamefont {A.~M.}\ \bibnamefont
  {Kadin}}\ and\ \bibinfo {author} {\bibfnamefont {S.~B.}\ \bibnamefont
  {Kaplan}},\ }\href {\doibase 10.1109/tasc.2017.2654858} {\bibfield  {journal}
  {\bibinfo  {journal} {{IEEE} Transactions on Applied Superconductivity}\
  }\textbf {\bibinfo {volume} {27}},\ \bibinfo {pages} {1} (\bibinfo {year}
  {2017})}\BibitemShut {NoStop}%
\bibitem [{\citenamefont {Pais}(2019)}]{SalvatorePais2019}%
  \BibitemOpen
  \bibfield  {author} {\bibinfo {author} {\bibfnamefont {S.~C.}\ \bibnamefont
  {Pais}},\ }\href
  {https://techlinkcenter.org/wp-content/uploads/2019/02/RTSC.pdf} {\enquote
  {\bibinfo {title} {Piezoelectricity-induced room temperature
  superconductor},}\ } (\bibinfo {year} {2019})\BibitemShut {NoStop}%
\bibitem [{\citenamefont {Jin}\ \emph {et~al.}(2003)\citenamefont {Jin},
  \citenamefont {Wang}, \citenamefont {Khachaturyan}, \citenamefont {Li},\ and\
  \citenamefont {Viehland}}]{Jin2003}%
  \BibitemOpen
  \bibfield  {author} {\bibinfo {author} {\bibfnamefont {Y.~M.}\ \bibnamefont
  {Jin}}, \bibinfo {author} {\bibfnamefont {Y.~U.}\ \bibnamefont {Wang}},
  \bibinfo {author} {\bibfnamefont {A.~G.}\ \bibnamefont {Khachaturyan}},
  \bibinfo {author} {\bibfnamefont {J.~F.}\ \bibnamefont {Li}}, \ and\ \bibinfo
  {author} {\bibfnamefont {D.}~\bibnamefont {Viehland}},\ }\href {\doibase
  10.1103/physrevlett.91.197601} {\bibfield  {journal} {\bibinfo  {journal}
  {Physical Review Letters}\ }\textbf {\bibinfo {volume} {91}} (\bibinfo {year}
  {2003}),\ 10.1103/physrevlett.91.197601}\BibitemShut {NoStop}%
\bibitem [{\citenamefont {Agar}\ \emph {et~al.}(2014)\citenamefont {Agar},
  \citenamefont {Mangalam}, \citenamefont {Damodaran}, \citenamefont {Velarde},
  \citenamefont {Karthik}, \citenamefont {Okatan}, \citenamefont {Chen},
  \citenamefont {Jesse}, \citenamefont {Balke}, \citenamefont {Kalinin},\ and\
  \citenamefont {Martin}}]{Agar2014}%
  \BibitemOpen
  \bibfield  {author} {\bibinfo {author} {\bibfnamefont {J.~C.}\ \bibnamefont
  {Agar}}, \bibinfo {author} {\bibfnamefont {R.~V.~K.}\ \bibnamefont
  {Mangalam}}, \bibinfo {author} {\bibfnamefont {A.~R.}\ \bibnamefont
  {Damodaran}}, \bibinfo {author} {\bibfnamefont {G.}~\bibnamefont {Velarde}},
  \bibinfo {author} {\bibfnamefont {J.}~\bibnamefont {Karthik}}, \bibinfo
  {author} {\bibfnamefont {M.~B.}\ \bibnamefont {Okatan}}, \bibinfo {author}
  {\bibfnamefont {Z.~H.}\ \bibnamefont {Chen}}, \bibinfo {author}
  {\bibfnamefont {S.}~\bibnamefont {Jesse}}, \bibinfo {author} {\bibfnamefont
  {N.}~\bibnamefont {Balke}}, \bibinfo {author} {\bibfnamefont {S.~V.}\
  \bibnamefont {Kalinin}}, \ and\ \bibinfo {author} {\bibfnamefont {L.~W.}\
  \bibnamefont {Martin}},\ }\href {\doibase 10.1002/admi.201400098} {\bibfield
  {journal} {\bibinfo  {journal} {Advanced Materials Interfaces}\ }\textbf
  {\bibinfo {volume} {1}},\ \bibinfo {pages} {1400098} (\bibinfo {year}
  {2014})}\BibitemShut {NoStop}%
\bibitem [{\citenamefont {Noheda}\ \emph {et~al.}(1999)\citenamefont {Noheda},
  \citenamefont {Cox}, \citenamefont {Shirane}, \citenamefont {Gonzalo},
  \citenamefont {Cross},\ and\ \citenamefont {Park}}]{Noheda1999}%
  \BibitemOpen
  \bibfield  {author} {\bibinfo {author} {\bibfnamefont {B.}~\bibnamefont
  {Noheda}}, \bibinfo {author} {\bibfnamefont {D.~E.}\ \bibnamefont {Cox}},
  \bibinfo {author} {\bibfnamefont {G.}~\bibnamefont {Shirane}}, \bibinfo
  {author} {\bibfnamefont {J.~A.}\ \bibnamefont {Gonzalo}}, \bibinfo {author}
  {\bibfnamefont {L.~E.}\ \bibnamefont {Cross}}, \ and\ \bibinfo {author}
  {\bibfnamefont {S.-E.}\ \bibnamefont {Park}},\ }\href {\doibase
  10.1063/1.123756} {\bibfield  {journal} {\bibinfo  {journal} {Applied Physics
  Letters}\ }\textbf {\bibinfo {volume} {74}},\ \bibinfo {pages} {2059}
  (\bibinfo {year} {1999})}\BibitemShut {NoStop}%
\bibitem [{\citenamefont {Catalan}\ \emph {et~al.}(2011)\citenamefont
  {Catalan}, \citenamefont {Lubk}, \citenamefont {Vlooswijk}, \citenamefont
  {Snoeck}, \citenamefont {Magen}, \citenamefont {Janssens}, \citenamefont
  {Rispens}, \citenamefont {Rijnders}, \citenamefont {Blank},\ and\
  \citenamefont {Noheda}}]{Catalan2011}%
  \BibitemOpen
  \bibfield  {author} {\bibinfo {author} {\bibfnamefont {G.}~\bibnamefont
  {Catalan}}, \bibinfo {author} {\bibfnamefont {A.}~\bibnamefont {Lubk}},
  \bibinfo {author} {\bibfnamefont {A.~H.~G.}\ \bibnamefont {Vlooswijk}},
  \bibinfo {author} {\bibfnamefont {E.}~\bibnamefont {Snoeck}}, \bibinfo
  {author} {\bibfnamefont {C.}~\bibnamefont {Magen}}, \bibinfo {author}
  {\bibfnamefont {A.}~\bibnamefont {Janssens}}, \bibinfo {author}
  {\bibfnamefont {G.}~\bibnamefont {Rispens}}, \bibinfo {author} {\bibfnamefont
  {G.}~\bibnamefont {Rijnders}}, \bibinfo {author} {\bibfnamefont {D.~H.~A.}\
  \bibnamefont {Blank}}, \ and\ \bibinfo {author} {\bibfnamefont
  {B.}~\bibnamefont {Noheda}},\ }\href {\doibase 10.1038/nmat3141} {\bibfield
  {journal} {\bibinfo  {journal} {Nature Materials}\ }\textbf {\bibinfo
  {volume} {10}},\ \bibinfo {pages} {963} (\bibinfo {year} {2011})}\BibitemShut
  {NoStop}%
\bibitem [{\citenamefont {Nelmes}\ and\ \citenamefont
  {Kuhs}(1985)}]{Nelmes1985}%
  \BibitemOpen
  \bibfield  {author} {\bibinfo {author} {\bibfnamefont {R.}~\bibnamefont
  {Nelmes}}\ and\ \bibinfo {author} {\bibfnamefont {W.}~\bibnamefont {Kuhs}},\
  }\href {\doibase 10.1016/0038-1098(85)90595-2} {\bibfield  {journal}
  {\bibinfo  {journal} {Solid State Communications}\ }\textbf {\bibinfo
  {volume} {54}},\ \bibinfo {pages} {721} (\bibinfo {year} {1985})}\BibitemShut
  {NoStop}%
\bibitem [{\citenamefont {Fthenakis}\ and\ \citenamefont
  {Ponomareva}(2017)}]{Fthenakis2017}%
  \BibitemOpen
  \bibfield  {author} {\bibinfo {author} {\bibfnamefont {Z.~G.}\ \bibnamefont
  {Fthenakis}}\ and\ \bibinfo {author} {\bibfnamefont {I.}~\bibnamefont
  {Ponomareva}},\ }\href {\doibase 10.1103/physrevb.96.184110} {\bibfield
  {journal} {\bibinfo  {journal} {Physical Review B}\ }\textbf {\bibinfo
  {volume} {96}} (\bibinfo {year} {2017}),\
  10.1103/physrevb.96.184110}\BibitemShut {NoStop}%
\bibitem [{\citenamefont {Rabe}(2013)}]{rabe2013antiferroelectricity}%
  \BibitemOpen
  \bibfield  {author} {\bibinfo {author} {\bibfnamefont {K.~M.}\ \bibnamefont
  {Rabe}},\ }\href@noop {} {\bibfield  {journal} {\bibinfo  {journal}
  {Functional Metal Oxides}\ ,\ \bibinfo {pages} {221}} (\bibinfo {year}
  {2013})}\BibitemShut {NoStop}%
\bibitem [{\citenamefont {Tagantsev}\ \emph {et~al.}(2013)\citenamefont
  {Tagantsev}, \citenamefont {Vaideeswaran}, \citenamefont {Vakhrushev},
  \citenamefont {Filimonov}, \citenamefont {Burkovsky}, \citenamefont
  {Shaganov}, \citenamefont {Andronikova}, \citenamefont {Rudskoy},
  \citenamefont {Baron}, \citenamefont {Uchiyama}, \citenamefont {Chernyshov},
  \citenamefont {Bosak}, \citenamefont {Ujma}, \citenamefont {Roleder},
  \citenamefont {Majchrowski}, \citenamefont {Ko},\ and\ \citenamefont
  {Setter}}]{Tagantsev2013}%
  \BibitemOpen
  \bibfield  {author} {\bibinfo {author} {\bibfnamefont {A.~K.}\ \bibnamefont
  {Tagantsev}}, \bibinfo {author} {\bibfnamefont {K.}~\bibnamefont
  {Vaideeswaran}}, \bibinfo {author} {\bibfnamefont {S.~B.}\ \bibnamefont
  {Vakhrushev}}, \bibinfo {author} {\bibfnamefont {A.~V.}\ \bibnamefont
  {Filimonov}}, \bibinfo {author} {\bibfnamefont {R.~G.}\ \bibnamefont
  {Burkovsky}}, \bibinfo {author} {\bibfnamefont {A.}~\bibnamefont {Shaganov}},
  \bibinfo {author} {\bibfnamefont {D.}~\bibnamefont {Andronikova}}, \bibinfo
  {author} {\bibfnamefont {A.~I.}\ \bibnamefont {Rudskoy}}, \bibinfo {author}
  {\bibfnamefont {A.~Q.~R.}\ \bibnamefont {Baron}}, \bibinfo {author}
  {\bibfnamefont {H.}~\bibnamefont {Uchiyama}}, \bibinfo {author}
  {\bibfnamefont {D.}~\bibnamefont {Chernyshov}}, \bibinfo {author}
  {\bibfnamefont {A.}~\bibnamefont {Bosak}}, \bibinfo {author} {\bibfnamefont
  {Z.}~\bibnamefont {Ujma}}, \bibinfo {author} {\bibfnamefont {K.}~\bibnamefont
  {Roleder}}, \bibinfo {author} {\bibfnamefont {A.}~\bibnamefont
  {Majchrowski}}, \bibinfo {author} {\bibfnamefont {J.-H.}\ \bibnamefont {Ko}},
  \ and\ \bibinfo {author} {\bibfnamefont {N.}~\bibnamefont {Setter}},\ }\href
  {\doibase 10.1038/ncomms3229} {\bibfield  {journal} {\bibinfo  {journal}
  {Nature Communications}\ }\textbf {\bibinfo {volume} {4}} (\bibinfo {year}
  {2013}),\ 10.1038/ncomms3229}\BibitemShut {NoStop}%
\bibitem [{\citenamefont {Hlinka}\ \emph {et~al.}(2014)\citenamefont {Hlinka},
  \citenamefont {Ostapchuk}, \citenamefont {Buixaderas}, \citenamefont
  {Kadlec}, \citenamefont {Kuzel}, \citenamefont {Gregora}, \citenamefont
  {Kroupa}, \citenamefont {Savinov}, \citenamefont {Klic}, \citenamefont
  {Drahokoupil}, \citenamefont {Etxebarria},\ and\ \citenamefont
  {Dec}}]{Hlinka2014}%
  \BibitemOpen
  \bibfield  {author} {\bibinfo {author} {\bibfnamefont {J.}~\bibnamefont
  {Hlinka}}, \bibinfo {author} {\bibfnamefont {T.}~\bibnamefont {Ostapchuk}},
  \bibinfo {author} {\bibfnamefont {E.}~\bibnamefont {Buixaderas}}, \bibinfo
  {author} {\bibfnamefont {C.}~\bibnamefont {Kadlec}}, \bibinfo {author}
  {\bibfnamefont {P.}~\bibnamefont {Kuzel}}, \bibinfo {author} {\bibfnamefont
  {I.}~\bibnamefont {Gregora}}, \bibinfo {author} {\bibfnamefont
  {J.}~\bibnamefont {Kroupa}}, \bibinfo {author} {\bibfnamefont
  {M.}~\bibnamefont {Savinov}}, \bibinfo {author} {\bibfnamefont
  {A.}~\bibnamefont {Klic}}, \bibinfo {author} {\bibfnamefont {J.}~\bibnamefont
  {Drahokoupil}}, \bibinfo {author} {\bibfnamefont {I.}~\bibnamefont
  {Etxebarria}}, \ and\ \bibinfo {author} {\bibfnamefont {J.}~\bibnamefont
  {Dec}},\ }\href {\doibase 10.1103/physrevlett.112.197601} {\bibfield
  {journal} {\bibinfo  {journal} {Physical Review Letters}\ }\textbf {\bibinfo
  {volume} {112}} (\bibinfo {year} {2014}),\
  10.1103/physrevlett.112.197601}\BibitemShut {NoStop}%
\bibitem [{\citenamefont {Mani}\ \emph {et~al.}(2015)\citenamefont {Mani},
  \citenamefont {Lisenkov},\ and\ \citenamefont {Ponomareva}}]{Mani2015}%
  \BibitemOpen
  \bibfield  {author} {\bibinfo {author} {\bibfnamefont {B.~K.}\ \bibnamefont
  {Mani}}, \bibinfo {author} {\bibfnamefont {S.}~\bibnamefont {Lisenkov}}, \
  and\ \bibinfo {author} {\bibfnamefont {I.}~\bibnamefont {Ponomareva}},\
  }\href {\doibase 10.1103/physrevb.91.134112} {\bibfield  {journal} {\bibinfo
  {journal} {Physical Review B}\ }\textbf {\bibinfo {volume} {91}} (\bibinfo
  {year} {2015}),\ 10.1103/physrevb.91.134112}\BibitemShut {NoStop}%
\bibitem [{\citenamefont {Raman}\ and\ \citenamefont
  {Nedungadi}(1940)}]{RAMAN1940}%
  \BibitemOpen
  \bibfield  {author} {\bibinfo {author} {\bibfnamefont {C.~V.}\ \bibnamefont
  {Raman}}\ and\ \bibinfo {author} {\bibfnamefont {T.~M.~K.}\ \bibnamefont
  {Nedungadi}},\ }\href {\doibase 10.1038/145147a0} {\bibfield  {journal}
  {\bibinfo  {journal} {Nature}\ }\textbf {\bibinfo {volume} {145}},\ \bibinfo
  {pages} {147} (\bibinfo {year} {1940})}\BibitemShut {NoStop}%
\bibitem [{\citenamefont {Cochran}(1959)}]{Cochran1959}%
  \BibitemOpen
  \bibfield  {author} {\bibinfo {author} {\bibfnamefont {W.}~\bibnamefont
  {Cochran}},\ }\href {\doibase 10.1103/physrevlett.3.412} {\bibfield
  {journal} {\bibinfo  {journal} {Physical Review Letters}\ }\textbf {\bibinfo
  {volume} {3}},\ \bibinfo {pages} {412} (\bibinfo {year} {1959})}\BibitemShut
  {NoStop}%
\bibitem [{\citenamefont {Cochran}(1960)}]{Cochran1960}%
  \BibitemOpen
  \bibfield  {author} {\bibinfo {author} {\bibfnamefont {W.}~\bibnamefont
  {Cochran}},\ }\href {\doibase 10.1080/00018736000101229} {\bibfield
  {journal} {\bibinfo  {journal} {Advances in Physics}\ }\textbf {\bibinfo
  {volume} {9}},\ \bibinfo {pages} {387} (\bibinfo {year} {1960})}\BibitemShut
  {NoStop}%
\bibitem [{\citenamefont {Anderson}(1960)}]{anderson1960pw}%
  \BibitemOpen
  \bibfield  {author} {\bibinfo {author} {\bibfnamefont {P.}~\bibnamefont
  {Anderson}},\ }\href@noop {} {\bibfield  {journal} {\bibinfo  {journal} {Izv.
  Akad. Nauk SSSR}\ ,\ \bibinfo {pages} {290}} (\bibinfo {year}
  {1960})}\BibitemShut {NoStop}%
\bibitem [{\citenamefont {Sicron}\ \emph {et~al.}(1994)\citenamefont {Sicron},
  \citenamefont {Ravel}, \citenamefont {Yacoby}, \citenamefont {Stern},
  \citenamefont {Dogan},\ and\ \citenamefont {Rehr}}]{Sicron1994}%
  \BibitemOpen
  \bibfield  {author} {\bibinfo {author} {\bibfnamefont {N.}~\bibnamefont
  {Sicron}}, \bibinfo {author} {\bibfnamefont {B.}~\bibnamefont {Ravel}},
  \bibinfo {author} {\bibfnamefont {Y.}~\bibnamefont {Yacoby}}, \bibinfo
  {author} {\bibfnamefont {E.~A.}\ \bibnamefont {Stern}}, \bibinfo {author}
  {\bibfnamefont {F.}~\bibnamefont {Dogan}}, \ and\ \bibinfo {author}
  {\bibfnamefont {J.~J.}\ \bibnamefont {Rehr}},\ }\href {\doibase
  10.1103/physrevb.50.13168} {\bibfield  {journal} {\bibinfo  {journal}
  {Physical Review B}\ }\textbf {\bibinfo {volume} {50}},\ \bibinfo {pages}
  {13168} (\bibinfo {year} {1994})}\BibitemShut {NoStop}%
\bibitem [{\citenamefont {Marton}\ and\ \citenamefont
  {Els\"{a}sser}(2011)}]{Marton2011}%
  \BibitemOpen
  \bibfield  {author} {\bibinfo {author} {\bibfnamefont {P.}~\bibnamefont
  {Marton}}\ and\ \bibinfo {author} {\bibfnamefont {C.}~\bibnamefont
  {Els\"{a}sser}},\ }\href {\doibase 10.1002/pssb.201046598} {\bibfield
  {journal} {\bibinfo  {journal} {physica status solidi (b)}\ }\textbf
  {\bibinfo {volume} {248}},\ \bibinfo {pages} {2222} (\bibinfo {year}
  {2011})}\BibitemShut {NoStop}%
\bibitem [{\citenamefont {Blok}\ \emph {et~al.}(2011)\citenamefont {Blok},
  \citenamefont {Blank}, \citenamefont {Rijnders}, \citenamefont {Rabe},\ and\
  \citenamefont {Vanderbilt}}]{Blok2011}%
  \BibitemOpen
  \bibfield  {author} {\bibinfo {author} {\bibfnamefont {J.~L.}\ \bibnamefont
  {Blok}}, \bibinfo {author} {\bibfnamefont {D.~H.~A.}\ \bibnamefont {Blank}},
  \bibinfo {author} {\bibfnamefont {G.}~\bibnamefont {Rijnders}}, \bibinfo
  {author} {\bibfnamefont {K.~M.}\ \bibnamefont {Rabe}}, \ and\ \bibinfo
  {author} {\bibfnamefont {D.}~\bibnamefont {Vanderbilt}},\ }\href {\doibase
  10.1103/physrevb.84.205413} {\bibfield  {journal} {\bibinfo  {journal}
  {Physical Review B}\ }\textbf {\bibinfo {volume} {84}} (\bibinfo {year}
  {2011}),\ 10.1103/physrevb.84.205413}\BibitemShut {NoStop}%
\bibitem [{\citenamefont {Wu}\ and\ \citenamefont {Krakauer}(2003)}]{Wu2003}%
  \BibitemOpen
  \bibfield  {author} {\bibinfo {author} {\bibfnamefont {Z.}~\bibnamefont
  {Wu}}\ and\ \bibinfo {author} {\bibfnamefont {H.}~\bibnamefont {Krakauer}},\
  }\href {\doibase 10.1103/physrevb.68.014112} {\bibfield  {journal} {\bibinfo
  {journal} {Physical Review B}\ }\textbf {\bibinfo {volume} {68}} (\bibinfo
  {year} {2003}),\ 10.1103/physrevb.68.014112}\BibitemShut {NoStop}%
\bibitem [{\citenamefont {Kim}\ \emph {et~al.}(2013)\citenamefont {Kim},
  \citenamefont {Lee}, \citenamefont {Cho}, \citenamefont {Shim},\ and\
  \citenamefont {Kim}}]{Kim2013}%
  \BibitemOpen
  \bibfield  {author} {\bibinfo {author} {\bibfnamefont {S.}~\bibnamefont
  {Kim}}, \bibinfo {author} {\bibfnamefont {W.-J.}\ \bibnamefont {Lee}},
  \bibinfo {author} {\bibfnamefont {Y.-H.}\ \bibnamefont {Cho}}, \bibinfo
  {author} {\bibfnamefont {M.}~\bibnamefont {Shim}}, \ and\ \bibinfo {author}
  {\bibfnamefont {S.}~\bibnamefont {Kim}},\ }\href {\doibase
  10.7567/jjap.52.091101} {\bibfield  {journal} {\bibinfo  {journal} {Japanese
  Journal of Applied Physics}\ }\textbf {\bibinfo {volume} {52}},\ \bibinfo
  {pages} {091101} (\bibinfo {year} {2013})}\BibitemShut {NoStop}%
\bibitem [{\citenamefont {Grinberg}\ \emph {et~al.}(2004)\citenamefont
  {Grinberg}, \citenamefont {Cooper},\ and\ \citenamefont
  {Rappe}}]{Grinberg2004}%
  \BibitemOpen
  \bibfield  {author} {\bibinfo {author} {\bibfnamefont {I.}~\bibnamefont
  {Grinberg}}, \bibinfo {author} {\bibfnamefont {V.}~\bibnamefont {Cooper}}, \
  and\ \bibinfo {author} {\bibfnamefont {A.}~\bibnamefont {Rappe}},\ }\href
  {\doibase 10.1103/physrevb.69.144118} {\bibfield  {journal} {\bibinfo
  {journal} {Physical Review B}\ }\textbf {\bibinfo {volume} {69}} (\bibinfo
  {year} {2004}),\ 10.1103/physrevb.69.144118}\BibitemShut {NoStop}%
\bibitem [{\citenamefont {Bungaro}\ and\ \citenamefont
  {Rabe}(2002)}]{Bungaro2002}%
  \BibitemOpen
  \bibfield  {author} {\bibinfo {author} {\bibfnamefont {C.}~\bibnamefont
  {Bungaro}}\ and\ \bibinfo {author} {\bibfnamefont {K.~M.}\ \bibnamefont
  {Rabe}},\ }\href {\doibase 10.1103/physrevb.65.224106} {\bibfield  {journal}
  {\bibinfo  {journal} {Physical Review B}\ }\textbf {\bibinfo {volume} {65}}
  (\bibinfo {year} {2002}),\ 10.1103/physrevb.65.224106}\BibitemShut {NoStop}%
\bibitem [{\citenamefont {Bellaiche}\ and\ \citenamefont
  {Vanderbilt}(2000)}]{Bellaiche2000}%
  \BibitemOpen
  \bibfield  {author} {\bibinfo {author} {\bibfnamefont {L.}~\bibnamefont
  {Bellaiche}}\ and\ \bibinfo {author} {\bibfnamefont {D.}~\bibnamefont
  {Vanderbilt}},\ }\href {\doibase 10.1103/physrevb.61.7877} {\bibfield
  {journal} {\bibinfo  {journal} {Physical Review B}\ }\textbf {\bibinfo
  {volume} {61}},\ \bibinfo {pages} {7877} (\bibinfo {year}
  {2000})}\BibitemShut {NoStop}%
\bibitem [{\citenamefont {Ramer}\ and\ \citenamefont
  {Rappe}(2000)}]{Ramer2000}%
  \BibitemOpen
  \bibfield  {author} {\bibinfo {author} {\bibfnamefont {N.~J.}\ \bibnamefont
  {Ramer}}\ and\ \bibinfo {author} {\bibfnamefont {A.~M.}\ \bibnamefont
  {Rappe}},\ }\href {\doibase 10.1103/physrevb.62.r743} {\bibfield  {journal}
  {\bibinfo  {journal} {Physical Review B}\ }\textbf {\bibinfo {volume} {62}},\
  \bibinfo {pages} {R743} (\bibinfo {year} {2000})}\BibitemShut {NoStop}%
\bibitem [{\citenamefont {Liu}\ \emph {et~al.}(2013)\citenamefont {Liu},
  \citenamefont {Shao}, \citenamefont {Yu}, \citenamefont {L\"{u}},
  \citenamefont {Li}, \citenamefont {Li},\ and\ \citenamefont {Cao}}]{Liu2013}%
  \BibitemOpen
  \bibfield  {author} {\bibinfo {author} {\bibfnamefont {S.-Y.}\ \bibnamefont
  {Liu}}, \bibinfo {author} {\bibfnamefont {Q.-S.}\ \bibnamefont {Shao}},
  \bibinfo {author} {\bibfnamefont {D.-S.}\ \bibnamefont {Yu}}, \bibinfo
  {author} {\bibfnamefont {Y.-K.}\ \bibnamefont {L\"{u}}}, \bibinfo {author}
  {\bibfnamefont {D.-J.}\ \bibnamefont {Li}}, \bibinfo {author} {\bibfnamefont
  {Y.}~\bibnamefont {Li}}, \ and\ \bibinfo {author} {\bibfnamefont {M.-S.}\
  \bibnamefont {Cao}},\ }\href {\doibase 10.1088/1674-1056/22/1/017702}
  {\bibfield  {journal} {\bibinfo  {journal} {Chinese Physics B}\ }\textbf
  {\bibinfo {volume} {22}},\ \bibinfo {pages} {017702} (\bibinfo {year}
  {2013})}\BibitemShut {NoStop}%
\bibitem [{\citenamefont {Bell}(2006)}]{Bell2006}%
  \BibitemOpen
  \bibfield  {author} {\bibinfo {author} {\bibfnamefont {A.~J.}\ \bibnamefont
  {Bell}},\ }\href {\doibase 10.1007/s10853-005-5913-9} {\bibfield  {journal}
  {\bibinfo  {journal} {Journal of Materials Science}\ }\textbf {\bibinfo
  {volume} {41}},\ \bibinfo {pages} {13} (\bibinfo {year} {2006})}\BibitemShut
  {NoStop}%
\bibitem [{\citenamefont {Bogdanov}\ \emph {et~al.}(2016)\citenamefont
  {Bogdanov}, \citenamefont {Mysovsky}, \citenamefont {Pickard},\ and\
  \citenamefont {Kimmel}}]{Bogdanov2016}%
  \BibitemOpen
  \bibfield  {author} {\bibinfo {author} {\bibfnamefont {A.}~\bibnamefont
  {Bogdanov}}, \bibinfo {author} {\bibfnamefont {A.}~\bibnamefont {Mysovsky}},
  \bibinfo {author} {\bibfnamefont {C.~J.}\ \bibnamefont {Pickard}}, \ and\
  \bibinfo {author} {\bibfnamefont {A.~V.}\ \bibnamefont {Kimmel}},\ }\href
  {\doibase 10.1039/c6cp04976a} {\bibfield  {journal} {\bibinfo  {journal}
  {Physical Chemistry Chemical Physics}\ }\textbf {\bibinfo {volume} {18}},\
  \bibinfo {pages} {28316} (\bibinfo {year} {2016})}\BibitemShut {NoStop}%
\bibitem [{\citenamefont {Glazer}(1972)}]{Glazer1972}%
  \BibitemOpen
  \bibfield  {author} {\bibinfo {author} {\bibfnamefont {A.~M.}\ \bibnamefont
  {Glazer}},\ }\href {\doibase 10.1107/s0567740872007976} {\bibfield  {journal}
  {\bibinfo  {journal} {Acta Crystallographica Section B Structural
  Crystallography and Crystal Chemistry}\ }\textbf {\bibinfo {volume} {28}},\
  \bibinfo {pages} {3384} (\bibinfo {year} {1972})}\BibitemShut {NoStop}%
\bibitem [{\citenamefont {Glazer}(1975)}]{Glazer1975}%
  \BibitemOpen
  \bibfield  {author} {\bibinfo {author} {\bibfnamefont {A.~M.}\ \bibnamefont
  {Glazer}},\ }\href {\doibase 10.1107/s0567739475001635} {\bibfield  {journal}
  {\bibinfo  {journal} {Acta Crystallographica Section A}\ }\textbf {\bibinfo
  {volume} {31}},\ \bibinfo {pages} {756} (\bibinfo {year} {1975})}\BibitemShut
  {NoStop}%
\bibitem [{\citenamefont {Gonze}\ \emph {et~al.}(2016)\citenamefont {Gonze},
  \citenamefont {Jollet}, \citenamefont {Araujo}, \citenamefont {Adams},
  \citenamefont {Amadon}, \citenamefont {Applencourt}, \citenamefont {Audouze},
  \citenamefont {Beuken}, \citenamefont {Bieder}, \citenamefont {Bokhanchuk},
  \citenamefont {Bousquet}, \citenamefont {Bruneval}, \citenamefont {Caliste},
  \citenamefont {C{\^{o}}t{\'{e}}}, \citenamefont {Dahm}, \citenamefont
  {Pieve}, \citenamefont {Delaveau}, \citenamefont {Gennaro}, \citenamefont
  {Dorado}, \citenamefont {Espejo}, \citenamefont {Geneste}, \citenamefont
  {Genovese}, \citenamefont {Gerossier}, \citenamefont {Giantomassi},
  \citenamefont {Gillet}, \citenamefont {Hamann}, \citenamefont {He},
  \citenamefont {Jomard}, \citenamefont {Janssen}, \citenamefont {Roux},
  \citenamefont {Levitt}, \citenamefont {Lherbier}, \citenamefont {Liu},
  \citenamefont {Luka{\v{c}}evi{\'{c}}}, \citenamefont {Martin}, \citenamefont
  {Martins}, \citenamefont {Oliveira}, \citenamefont {Ponc{\'{e}}},
  \citenamefont {Pouillon}, \citenamefont {Rangel}, \citenamefont {Rignanese},
  \citenamefont {Romero}, \citenamefont {Rousseau}, \citenamefont {Rubel},
  \citenamefont {Shukri}, \citenamefont {Stankovski}, \citenamefont {Torrent},
  \citenamefont {Setten}, \citenamefont {Troeye}, \citenamefont {Verstraete},
  \citenamefont {Waroquiers}, \citenamefont {Wiktor}, \citenamefont {Xu},
  \citenamefont {Zhou},\ and\ \citenamefont {Zwanziger}}]{Gonze2016}%
  \BibitemOpen
  \bibfield  {author} {\bibinfo {author} {\bibfnamefont {X.}~\bibnamefont
  {Gonze}}, \bibinfo {author} {\bibfnamefont {F.}~\bibnamefont {Jollet}},
  \bibinfo {author} {\bibfnamefont {F.~A.}\ \bibnamefont {Araujo}}, \bibinfo
  {author} {\bibfnamefont {D.}~\bibnamefont {Adams}}, \bibinfo {author}
  {\bibfnamefont {B.}~\bibnamefont {Amadon}}, \bibinfo {author} {\bibfnamefont
  {T.}~\bibnamefont {Applencourt}}, \bibinfo {author} {\bibfnamefont
  {C.}~\bibnamefont {Audouze}}, \bibinfo {author} {\bibfnamefont {J.-M.}\
  \bibnamefont {Beuken}}, \bibinfo {author} {\bibfnamefont {J.}~\bibnamefont
  {Bieder}}, \bibinfo {author} {\bibfnamefont {A.}~\bibnamefont {Bokhanchuk}},
  \bibinfo {author} {\bibfnamefont {E.}~\bibnamefont {Bousquet}}, \bibinfo
  {author} {\bibfnamefont {F.}~\bibnamefont {Bruneval}}, \bibinfo {author}
  {\bibfnamefont {D.}~\bibnamefont {Caliste}}, \bibinfo {author} {\bibfnamefont
  {M.}~\bibnamefont {C{\^{o}}t{\'{e}}}}, \bibinfo {author} {\bibfnamefont
  {F.}~\bibnamefont {Dahm}}, \bibinfo {author} {\bibfnamefont {F.~D.}\
  \bibnamefont {Pieve}}, \bibinfo {author} {\bibfnamefont {M.}~\bibnamefont
  {Delaveau}}, \bibinfo {author} {\bibfnamefont {M.~D.}\ \bibnamefont
  {Gennaro}}, \bibinfo {author} {\bibfnamefont {B.}~\bibnamefont {Dorado}},
  \bibinfo {author} {\bibfnamefont {C.}~\bibnamefont {Espejo}}, \bibinfo
  {author} {\bibfnamefont {G.}~\bibnamefont {Geneste}}, \bibinfo {author}
  {\bibfnamefont {L.}~\bibnamefont {Genovese}}, \bibinfo {author}
  {\bibfnamefont {A.}~\bibnamefont {Gerossier}}, \bibinfo {author}
  {\bibfnamefont {M.}~\bibnamefont {Giantomassi}}, \bibinfo {author}
  {\bibfnamefont {Y.}~\bibnamefont {Gillet}}, \bibinfo {author} {\bibfnamefont
  {D.}~\bibnamefont {Hamann}}, \bibinfo {author} {\bibfnamefont
  {L.}~\bibnamefont {He}}, \bibinfo {author} {\bibfnamefont {G.}~\bibnamefont
  {Jomard}}, \bibinfo {author} {\bibfnamefont {J.~L.}\ \bibnamefont {Janssen}},
  \bibinfo {author} {\bibfnamefont {S.~L.}\ \bibnamefont {Roux}}, \bibinfo
  {author} {\bibfnamefont {A.}~\bibnamefont {Levitt}}, \bibinfo {author}
  {\bibfnamefont {A.}~\bibnamefont {Lherbier}}, \bibinfo {author}
  {\bibfnamefont {F.}~\bibnamefont {Liu}}, \bibinfo {author} {\bibfnamefont
  {I.}~\bibnamefont {Luka{\v{c}}evi{\'{c}}}}, \bibinfo {author} {\bibfnamefont
  {A.}~\bibnamefont {Martin}}, \bibinfo {author} {\bibfnamefont
  {C.}~\bibnamefont {Martins}}, \bibinfo {author} {\bibfnamefont
  {M.}~\bibnamefont {Oliveira}}, \bibinfo {author} {\bibfnamefont
  {S.}~\bibnamefont {Ponc{\'{e}}}}, \bibinfo {author} {\bibfnamefont
  {Y.}~\bibnamefont {Pouillon}}, \bibinfo {author} {\bibfnamefont
  {T.}~\bibnamefont {Rangel}}, \bibinfo {author} {\bibfnamefont {G.-M.}\
  \bibnamefont {Rignanese}}, \bibinfo {author} {\bibfnamefont {A.}~\bibnamefont
  {Romero}}, \bibinfo {author} {\bibfnamefont {B.}~\bibnamefont {Rousseau}},
  \bibinfo {author} {\bibfnamefont {O.}~\bibnamefont {Rubel}}, \bibinfo
  {author} {\bibfnamefont {A.}~\bibnamefont {Shukri}}, \bibinfo {author}
  {\bibfnamefont {M.}~\bibnamefont {Stankovski}}, \bibinfo {author}
  {\bibfnamefont {M.}~\bibnamefont {Torrent}}, \bibinfo {author} {\bibfnamefont
  {M.~V.}\ \bibnamefont {Setten}}, \bibinfo {author} {\bibfnamefont {B.~V.}\
  \bibnamefont {Troeye}}, \bibinfo {author} {\bibfnamefont {M.}~\bibnamefont
  {Verstraete}}, \bibinfo {author} {\bibfnamefont {D.}~\bibnamefont
  {Waroquiers}}, \bibinfo {author} {\bibfnamefont {J.}~\bibnamefont {Wiktor}},
  \bibinfo {author} {\bibfnamefont {B.}~\bibnamefont {Xu}}, \bibinfo {author}
  {\bibfnamefont {A.}~\bibnamefont {Zhou}}, \ and\ \bibinfo {author}
  {\bibfnamefont {J.}~\bibnamefont {Zwanziger}},\ }\href {\doibase
  10.1016/j.cpc.2016.04.003} {\bibfield  {journal} {\bibinfo  {journal}
  {Computer Physics Communications}\ }\textbf {\bibinfo {volume} {205}},\
  \bibinfo {pages} {106} (\bibinfo {year} {2016})}\BibitemShut {NoStop}%
\bibitem [{\citenamefont {Gonze}\ \emph {et~al.}(2009)\citenamefont {Gonze},
  \citenamefont {Amadon}, \citenamefont {Anglade}, \citenamefont {Beuken},
  \citenamefont {Bottin}, \citenamefont {Boulanger}, \citenamefont {Bruneval},
  \citenamefont {Caliste}, \citenamefont {Caracas}, \citenamefont
  {C{\^{o}}t{\'{e}}}, \citenamefont {Deutsch}, \citenamefont {Genovese},
  \citenamefont {Ghosez}, \citenamefont {Giantomassi}, \citenamefont
  {Goedecker}, \citenamefont {Hamann}, \citenamefont {Hermet}, \citenamefont
  {Jollet}, \citenamefont {Jomard}, \citenamefont {Leroux}, \citenamefont
  {Mancini}, \citenamefont {Mazevet}, \citenamefont {Oliveira}, \citenamefont
  {Onida}, \citenamefont {Pouillon}, \citenamefont {Rangel}, \citenamefont
  {Rignanese}, \citenamefont {Sangalli}, \citenamefont {Shaltaf}, \citenamefont
  {Torrent}, \citenamefont {Verstraete}, \citenamefont {Zerah},\ and\
  \citenamefont {Zwanziger}}]{Gonze2009}%
  \BibitemOpen
  \bibfield  {author} {\bibinfo {author} {\bibfnamefont {X.}~\bibnamefont
  {Gonze}}, \bibinfo {author} {\bibfnamefont {B.}~\bibnamefont {Amadon}},
  \bibinfo {author} {\bibfnamefont {P.-M.}\ \bibnamefont {Anglade}}, \bibinfo
  {author} {\bibfnamefont {J.-M.}\ \bibnamefont {Beuken}}, \bibinfo {author}
  {\bibfnamefont {F.}~\bibnamefont {Bottin}}, \bibinfo {author} {\bibfnamefont
  {P.}~\bibnamefont {Boulanger}}, \bibinfo {author} {\bibfnamefont
  {F.}~\bibnamefont {Bruneval}}, \bibinfo {author} {\bibfnamefont
  {D.}~\bibnamefont {Caliste}}, \bibinfo {author} {\bibfnamefont
  {R.}~\bibnamefont {Caracas}}, \bibinfo {author} {\bibfnamefont
  {M.}~\bibnamefont {C{\^{o}}t{\'{e}}}}, \bibinfo {author} {\bibfnamefont
  {T.}~\bibnamefont {Deutsch}}, \bibinfo {author} {\bibfnamefont
  {L.}~\bibnamefont {Genovese}}, \bibinfo {author} {\bibfnamefont
  {P.}~\bibnamefont {Ghosez}}, \bibinfo {author} {\bibfnamefont
  {M.}~\bibnamefont {Giantomassi}}, \bibinfo {author} {\bibfnamefont
  {S.}~\bibnamefont {Goedecker}}, \bibinfo {author} {\bibfnamefont
  {D.}~\bibnamefont {Hamann}}, \bibinfo {author} {\bibfnamefont
  {P.}~\bibnamefont {Hermet}}, \bibinfo {author} {\bibfnamefont
  {F.}~\bibnamefont {Jollet}}, \bibinfo {author} {\bibfnamefont
  {G.}~\bibnamefont {Jomard}}, \bibinfo {author} {\bibfnamefont
  {S.}~\bibnamefont {Leroux}}, \bibinfo {author} {\bibfnamefont
  {M.}~\bibnamefont {Mancini}}, \bibinfo {author} {\bibfnamefont
  {S.}~\bibnamefont {Mazevet}}, \bibinfo {author} {\bibfnamefont
  {M.}~\bibnamefont {Oliveira}}, \bibinfo {author} {\bibfnamefont
  {G.}~\bibnamefont {Onida}}, \bibinfo {author} {\bibfnamefont
  {Y.}~\bibnamefont {Pouillon}}, \bibinfo {author} {\bibfnamefont
  {T.}~\bibnamefont {Rangel}}, \bibinfo {author} {\bibfnamefont {G.-M.}\
  \bibnamefont {Rignanese}}, \bibinfo {author} {\bibfnamefont {D.}~\bibnamefont
  {Sangalli}}, \bibinfo {author} {\bibfnamefont {R.}~\bibnamefont {Shaltaf}},
  \bibinfo {author} {\bibfnamefont {M.}~\bibnamefont {Torrent}}, \bibinfo
  {author} {\bibfnamefont {M.}~\bibnamefont {Verstraete}}, \bibinfo {author}
  {\bibfnamefont {G.}~\bibnamefont {Zerah}}, \ and\ \bibinfo {author}
  {\bibfnamefont {J.}~\bibnamefont {Zwanziger}},\ }\href {\doibase
  10.1016/j.cpc.2009.07.007} {\bibfield  {journal} {\bibinfo  {journal}
  {Computer Physics Communications}\ }\textbf {\bibinfo {volume} {180}},\
  \bibinfo {pages} {2582} (\bibinfo {year} {2009})}\BibitemShut {NoStop}%
\bibitem [{\citenamefont {Hamann}(2013)}]{hamann2013optimized}%
  \BibitemOpen
  \bibfield  {author} {\bibinfo {author} {\bibfnamefont {D.~R.}\ \bibnamefont
  {Hamann}},\ }\href {https://doi.org/10.1103/physrevb.88.085117} {\bibfield
  {journal} {\bibinfo  {journal} {Phys. Rev. B}\ }\textbf {\bibinfo {volume}
  {88}} (\bibinfo {year} {2013})}\BibitemShut {NoStop}%
\bibitem [{\citenamefont {van Setten}\ \emph {et~al.}(2018)\citenamefont {van
  Setten}, \citenamefont {Giantomassi}, \citenamefont {Bousquet}, \citenamefont
  {Verstraete}, \citenamefont {Hamann}, \citenamefont {Gonze},\ and\
  \citenamefont {Rignanese}}]{van2018pseudodojo}%
  \BibitemOpen
  \bibfield  {author} {\bibinfo {author} {\bibfnamefont {M.}~\bibnamefont {van
  Setten}}, \bibinfo {author} {\bibfnamefont {M.}~\bibnamefont {Giantomassi}},
  \bibinfo {author} {\bibfnamefont {E.}~\bibnamefont {Bousquet}}, \bibinfo
  {author} {\bibfnamefont {M.}~\bibnamefont {Verstraete}}, \bibinfo {author}
  {\bibfnamefont {D.}~\bibnamefont {Hamann}}, \bibinfo {author} {\bibfnamefont
  {X.}~\bibnamefont {Gonze}}, \ and\ \bibinfo {author} {\bibfnamefont {G.-M.}\
  \bibnamefont {Rignanese}},\ }\href
  {https://doi.org/10.1016/j.cpc.2018.01.012} {\bibfield  {journal} {\bibinfo
  {journal} {Comput. Phys. Commun.}\ }\textbf {\bibinfo {volume} {226}},\
  \bibinfo {pages} {39} (\bibinfo {year} {2018})}\BibitemShut {NoStop}%
\bibitem [{\citenamefont {Monkhorst}\ and\ \citenamefont
  {Pack}(1976)}]{monkhorst1976special}%
  \BibitemOpen
  \bibfield  {author} {\bibinfo {author} {\bibfnamefont {H.~J.}\ \bibnamefont
  {Monkhorst}}\ and\ \bibinfo {author} {\bibfnamefont {J.~D.}\ \bibnamefont
  {Pack}},\ }\href {https://doi.org/10.1103/physrevb.13.5188} {\bibfield
  {journal} {\bibinfo  {journal} {Phys. Rev. B}\ }\textbf {\bibinfo {volume}
  {13}},\ \bibinfo {pages} {5188} (\bibinfo {year} {1976})}\BibitemShut
  {NoStop}%
\bibitem [{\citenamefont {Perdew}\ \emph {et~al.}(2008)\citenamefont {Perdew},
  \citenamefont {Ruzsinszky}, \citenamefont {Csonka}, \citenamefont {Vydrov},
  \citenamefont {Scuseria}, \citenamefont {Constantin}, \citenamefont {Zhou},\
  and\ \citenamefont {Burke}}]{Perdew2008}%
  \BibitemOpen
  \bibfield  {author} {\bibinfo {author} {\bibfnamefont {J.~P.}\ \bibnamefont
  {Perdew}}, \bibinfo {author} {\bibfnamefont {A.}~\bibnamefont {Ruzsinszky}},
  \bibinfo {author} {\bibfnamefont {G.~I.}\ \bibnamefont {Csonka}}, \bibinfo
  {author} {\bibfnamefont {O.~A.}\ \bibnamefont {Vydrov}}, \bibinfo {author}
  {\bibfnamefont {G.~E.}\ \bibnamefont {Scuseria}}, \bibinfo {author}
  {\bibfnamefont {L.~A.}\ \bibnamefont {Constantin}}, \bibinfo {author}
  {\bibfnamefont {X.}~\bibnamefont {Zhou}}, \ and\ \bibinfo {author}
  {\bibfnamefont {K.}~\bibnamefont {Burke}},\ }\href {\doibase
  10.1103/physrevlett.100.136406} {\bibfield  {journal} {\bibinfo  {journal}
  {Physical Review Letters}\ }\textbf {\bibinfo {volume} {100}} (\bibinfo
  {year} {2008}),\ 10.1103/physrevlett.100.136406}\BibitemShut {NoStop}%
\bibitem [{\citenamefont {Marques}\ \emph {et~al.}(2012)\citenamefont
  {Marques}, \citenamefont {Oliveira},\ and\ \citenamefont
  {Burnus}}]{Marques2012}%
  \BibitemOpen
  \bibfield  {author} {\bibinfo {author} {\bibfnamefont {M.~A.}\ \bibnamefont
  {Marques}}, \bibinfo {author} {\bibfnamefont {M.~J.}\ \bibnamefont
  {Oliveira}}, \ and\ \bibinfo {author} {\bibfnamefont {T.}~\bibnamefont
  {Burnus}},\ }\href {\doibase 10.1016/j.cpc.2012.05.007} {\bibfield  {journal}
  {\bibinfo  {journal} {Computer Physics Communications}\ }\textbf {\bibinfo
  {volume} {183}},\ \bibinfo {pages} {2272} (\bibinfo {year}
  {2012})}\BibitemShut {NoStop}%
\bibitem [{\citenamefont {Zhang}\ \emph {et~al.}(2017)\citenamefont {Zhang},
  \citenamefont {Sun}, \citenamefont {Perdew},\ and\ \citenamefont
  {Wu}}]{Zhang2017}%
  \BibitemOpen
  \bibfield  {author} {\bibinfo {author} {\bibfnamefont {Y.}~\bibnamefont
  {Zhang}}, \bibinfo {author} {\bibfnamefont {J.}~\bibnamefont {Sun}}, \bibinfo
  {author} {\bibfnamefont {J.~P.}\ \bibnamefont {Perdew}}, \ and\ \bibinfo
  {author} {\bibfnamefont {X.}~\bibnamefont {Wu}},\ }\href {\doibase
  10.1103/physrevb.96.035143} {\bibfield  {journal} {\bibinfo  {journal}
  {Physical Review B}\ }\textbf {\bibinfo {volume} {96}} (\bibinfo {year}
  {2017}),\ 10.1103/physrevb.96.035143}\BibitemShut {NoStop}%
\bibitem [{\citenamefont {Mabud}\ and\ \citenamefont
  {Glazer}(1979)}]{Mabud1979}%
  \BibitemOpen
  \bibfield  {author} {\bibinfo {author} {\bibfnamefont {S.~A.}\ \bibnamefont
  {Mabud}}\ and\ \bibinfo {author} {\bibfnamefont {A.~M.}\ \bibnamefont
  {Glazer}},\ }\href {\doibase 10.1107/s0021889879011754} {\bibfield  {journal}
  {\bibinfo  {journal} {Journal of Applied Crystallography}\ }\textbf {\bibinfo
  {volume} {12}},\ \bibinfo {pages} {49} (\bibinfo {year} {1979})}\BibitemShut
  {NoStop}%
\bibitem [{\citenamefont {Sawaguchi}(1953)}]{Sawaguchi1953}%
  \BibitemOpen
  \bibfield  {author} {\bibinfo {author} {\bibfnamefont {E.}~\bibnamefont
  {Sawaguchi}},\ }\href {\doibase 10.1143/jpsj.8.615} {\bibfield  {journal}
  {\bibinfo  {journal} {Journal of the Physical Society of Japan}\ }\textbf
  {\bibinfo {volume} {8}},\ \bibinfo {pages} {615} (\bibinfo {year}
  {1953})}\BibitemShut {NoStop}%
\bibitem [{\citenamefont {Gonze}\ and\ \citenamefont {Lee}(1997)}]{Gonze1997}%
  \BibitemOpen
  \bibfield  {author} {\bibinfo {author} {\bibfnamefont {X.}~\bibnamefont
  {Gonze}}\ and\ \bibinfo {author} {\bibfnamefont {C.}~\bibnamefont {Lee}},\
  }\href {\doibase 10.1103/physrevb.55.10355} {\bibfield  {journal} {\bibinfo
  {journal} {Physical Review B}\ }\textbf {\bibinfo {volume} {55}},\ \bibinfo
  {pages} {10355} (\bibinfo {year} {1997})}\BibitemShut {NoStop}%
\bibitem [{\citenamefont {Baroni}\ \emph {et~al.}(2001)\citenamefont {Baroni},
  \citenamefont {de~Gironcoli}, \citenamefont {Corso},\ and\ \citenamefont
  {Giannozzi}}]{Baroni2001}%
  \BibitemOpen
  \bibfield  {author} {\bibinfo {author} {\bibfnamefont {S.}~\bibnamefont
  {Baroni}}, \bibinfo {author} {\bibfnamefont {S.}~\bibnamefont
  {de~Gironcoli}}, \bibinfo {author} {\bibfnamefont {A.~D.}\ \bibnamefont
  {Corso}}, \ and\ \bibinfo {author} {\bibfnamefont {P.}~\bibnamefont
  {Giannozzi}},\ }\href {\doibase 10.1103/revmodphys.73.515} {\bibfield
  {journal} {\bibinfo  {journal} {Reviews of Modern Physics}\ }\textbf
  {\bibinfo {volume} {73}},\ \bibinfo {pages} {515} (\bibinfo {year}
  {2001})}\BibitemShut {NoStop}%
\bibitem [{\citenamefont {Zhong}\ \emph {et~al.}(1994)\citenamefont {Zhong},
  \citenamefont {King-Smith},\ and\ \citenamefont {Vanderbilt}}]{Zhong1994}%
  \BibitemOpen
  \bibfield  {author} {\bibinfo {author} {\bibfnamefont {W.}~\bibnamefont
  {Zhong}}, \bibinfo {author} {\bibfnamefont {R.~D.}\ \bibnamefont
  {King-Smith}}, \ and\ \bibinfo {author} {\bibfnamefont {D.}~\bibnamefont
  {Vanderbilt}},\ }\href {\doibase 10.1103/physrevlett.72.3618} {\bibfield
  {journal} {\bibinfo  {journal} {Physical Review Letters}\ }\textbf {\bibinfo
  {volume} {72}},\ \bibinfo {pages} {3618} (\bibinfo {year}
  {1994})}\BibitemShut {NoStop}%
\bibitem [{\citenamefont {Henry}\ and\ \citenamefont
  {Hopfield}(1965)}]{Henry1965}%
  \BibitemOpen
  \bibfield  {author} {\bibinfo {author} {\bibfnamefont {C.~H.}\ \bibnamefont
  {Henry}}\ and\ \bibinfo {author} {\bibfnamefont {J.~J.}\ \bibnamefont
  {Hopfield}},\ }\href {\doibase 10.1103/physrevlett.15.964} {\bibfield
  {journal} {\bibinfo  {journal} {Physical Review Letters}\ }\textbf {\bibinfo
  {volume} {15}},\ \bibinfo {pages} {964} (\bibinfo {year} {1965})}\BibitemShut
  {NoStop}%
\bibitem [{\citenamefont {Gonze}(1997)}]{Gonze1997-2}%
  \BibitemOpen
  \bibfield  {author} {\bibinfo {author} {\bibfnamefont {X.}~\bibnamefont
  {Gonze}},\ }\href {\doibase 10.1103/physrevb.55.10337} {\bibfield  {journal}
  {\bibinfo  {journal} {Physical Review B}\ }\textbf {\bibinfo {volume} {55}},\
  \bibinfo {pages} {10337} (\bibinfo {year} {1997})}\BibitemShut {NoStop}%
\bibitem [{\citenamefont {Ghosez}(2000)}]{Ghosez2000}%
  \BibitemOpen
  \bibfield  {author} {\bibinfo {author} {\bibfnamefont {P.}~\bibnamefont
  {Ghosez}},\ }in\ \href {\doibase 10.1063/1.1324445} {\emph {\bibinfo
  {booktitle} {{AIP} Conference Proceedings}}}\ (\bibinfo  {publisher}
  {{AIP}},\ \bibinfo {year} {2000})\BibitemShut {NoStop}%
\bibitem [{\citenamefont {Togo}\ and\ \citenamefont {Tanaka}(2015)}]{Togo2015}%
  \BibitemOpen
  \bibfield  {author} {\bibinfo {author} {\bibfnamefont {A.}~\bibnamefont
  {Togo}}\ and\ \bibinfo {author} {\bibfnamefont {I.}~\bibnamefont {Tanaka}},\
  }\href {\doibase 10.1016/j.scriptamat.2015.07.021} {\bibfield  {journal}
  {\bibinfo  {journal} {Scripta Materialia}\ }\textbf {\bibinfo {volume}
  {108}},\ \bibinfo {pages} {1} (\bibinfo {year} {2015})}\BibitemShut {NoStop}%
\bibitem [{not()}]{note:SM}%
  \BibitemOpen
  \href@noop {} {}\bibinfo {note} {See Supplemental Material at [URL to be
  inserted by publisher] for a full tabulation of the soft modes for each
  calculation, a demonstration of the equivalence of DFPT to the FDM and the
  full phonon dispersion curves (including real branches) for each
  calculation.}\BibitemShut {Stop}%
\bibitem [{\citenamefont {Vegard}(1921)}]{Vegard1921}%
  \BibitemOpen
  \bibfield  {author} {\bibinfo {author} {\bibfnamefont {L.}~\bibnamefont
  {Vegard}},\ }\href {\doibase 10.1007/bf01349680} {\bibfield  {journal}
  {\bibinfo  {journal} {Zeitschrift fur Physik}\ }\textbf {\bibinfo {volume}
  {5}},\ \bibinfo {pages} {17} (\bibinfo {year} {1921})}\BibitemShut {NoStop}%
\bibitem [{\citenamefont {Bl\"{o}chl}\ \emph {et~al.}(1994)\citenamefont
  {Bl\"{o}chl}, \citenamefont {Jepsen},\ and\ \citenamefont
  {Andersen}}]{Blchl1994}%
  \BibitemOpen
  \bibfield  {author} {\bibinfo {author} {\bibfnamefont {P.~E.}\ \bibnamefont
  {Bl\"{o}chl}}, \bibinfo {author} {\bibfnamefont {O.}~\bibnamefont {Jepsen}},
  \ and\ \bibinfo {author} {\bibfnamefont {O.~K.}\ \bibnamefont {Andersen}},\
  }\href {\doibase 10.1103/physrevb.49.16223} {\bibfield  {journal} {\bibinfo
  {journal} {Physical Review B}\ }\textbf {\bibinfo {volume} {49}},\ \bibinfo
  {pages} {16223} (\bibinfo {year} {1994})}\BibitemShut {NoStop}%
\bibitem [{\citenamefont {Stokes}\ and\ \citenamefont
  {Hatch}(2005)}]{Stokes2005}%
  \BibitemOpen
  \bibfield  {author} {\bibinfo {author} {\bibfnamefont {H.~T.}\ \bibnamefont
  {Stokes}}\ and\ \bibinfo {author} {\bibfnamefont {D.~M.}\ \bibnamefont
  {Hatch}},\ }\href {\doibase 10.1107/s0021889804031528} {\bibfield  {journal}
  {\bibinfo  {journal} {Journal of Applied Crystallography}\ }\textbf {\bibinfo
  {volume} {38}},\ \bibinfo {pages} {237} (\bibinfo {year} {2005})}\BibitemShut
  {NoStop}%
\bibitem [{\citenamefont {Campbell}\ \emph {et~al.}(2006)\citenamefont
  {Campbell}, \citenamefont {Stokes}, \citenamefont {Tanner},\ and\
  \citenamefont {Hatch}}]{Campbell2006}%
  \BibitemOpen
  \bibfield  {author} {\bibinfo {author} {\bibfnamefont {B.~J.}\ \bibnamefont
  {Campbell}}, \bibinfo {author} {\bibfnamefont {H.~T.}\ \bibnamefont
  {Stokes}}, \bibinfo {author} {\bibfnamefont {D.~E.}\ \bibnamefont {Tanner}},
  \ and\ \bibinfo {author} {\bibfnamefont {D.~M.}\ \bibnamefont {Hatch}},\
  }\href {\doibase 10.1107/s0021889806014075} {\bibfield  {journal} {\bibinfo
  {journal} {Journal of Applied Crystallography}\ }\textbf {\bibinfo {volume}
  {39}},\ \bibinfo {pages} {607} (\bibinfo {year} {2006})}\BibitemShut
  {NoStop}%
\bibitem [{\citenamefont {Miranda}(2019)}]{PhononWeb}%
  \BibitemOpen
  \bibfield  {author} {\bibinfo {author} {\bibfnamefont {H.}~\bibnamefont
  {Miranda}},\ }\href {http://henriquemiranda.github.io/phononwebsite/}
  {\enquote {\bibinfo {title} {Phonon website},}\ } (\bibinfo {year}
  {2019})\BibitemShut {NoStop}%
\bibitem [{\citenamefont {Aroyo}\ \emph {et~al.}(2006)\citenamefont {Aroyo},
  \citenamefont {Perez-Mato}, \citenamefont {Capillas}, \citenamefont
  {Kroumova}, \citenamefont {Ivantchev}, \citenamefont {Madariaga},
  \citenamefont {Kirov},\ and\ \citenamefont {Wondratschek}}]{Aroyo2006}%
  \BibitemOpen
  \bibfield  {author} {\bibinfo {author} {\bibfnamefont {M.~I.}\ \bibnamefont
  {Aroyo}}, \bibinfo {author} {\bibfnamefont {J.~M.}\ \bibnamefont
  {Perez-Mato}}, \bibinfo {author} {\bibfnamefont {C.}~\bibnamefont
  {Capillas}}, \bibinfo {author} {\bibfnamefont {E.}~\bibnamefont {Kroumova}},
  \bibinfo {author} {\bibfnamefont {S.}~\bibnamefont {Ivantchev}}, \bibinfo
  {author} {\bibfnamefont {G.}~\bibnamefont {Madariaga}}, \bibinfo {author}
  {\bibfnamefont {A.}~\bibnamefont {Kirov}}, \ and\ \bibinfo {author}
  {\bibfnamefont {H.}~\bibnamefont {Wondratschek}},\ }\href {\doibase
  10.1524/zkri.2006.221.1.15} {\bibfield  {journal} {\bibinfo  {journal}
  {Zeitschrift f\"{u}r Kristallographie - Crystalline Materials}\ }\textbf
  {\bibinfo {volume} {221}} (\bibinfo {year} {2006}),\
  10.1524/zkri.2006.221.1.15}\BibitemShut {NoStop}%
\bibitem [{\citenamefont {Ghosez}\ \emph {et~al.}(1999)\citenamefont {Ghosez},
  \citenamefont {Cockayne}, \citenamefont {Waghmare},\ and\ \citenamefont
  {Rabe}}]{Ghosez1999}%
  \BibitemOpen
  \bibfield  {author} {\bibinfo {author} {\bibfnamefont {P.}~\bibnamefont
  {Ghosez}}, \bibinfo {author} {\bibfnamefont {E.}~\bibnamefont {Cockayne}},
  \bibinfo {author} {\bibfnamefont {U.~V.}\ \bibnamefont {Waghmare}}, \ and\
  \bibinfo {author} {\bibfnamefont {K.~M.}\ \bibnamefont {Rabe}},\ }\href
  {\doibase 10.1103/physrevb.60.836} {\bibfield  {journal} {\bibinfo  {journal}
  {Physical Review B}\ }\textbf {\bibinfo {volume} {60}},\ \bibinfo {pages}
  {836} (\bibinfo {year} {1999})}\BibitemShut {NoStop}%
\bibitem [{\citenamefont {{\'{I}}{\~{n}}iguez}\ \emph
  {et~al.}(2014)\citenamefont {{\'{I}}{\~{n}}iguez}, \citenamefont {Stengel},
  \citenamefont {Prosandeev},\ and\ \citenamefont {Bellaiche}}]{inguez2014}%
  \BibitemOpen
  \bibfield  {author} {\bibinfo {author} {\bibfnamefont {J.}~\bibnamefont
  {{\'{I}}{\~{n}}iguez}}, \bibinfo {author} {\bibfnamefont {M.}~\bibnamefont
  {Stengel}}, \bibinfo {author} {\bibfnamefont {S.}~\bibnamefont {Prosandeev}},
  \ and\ \bibinfo {author} {\bibfnamefont {L.}~\bibnamefont {Bellaiche}},\
  }\href {\doibase 10.1103/physrevb.90.220103} {\bibfield  {journal} {\bibinfo
  {journal} {Physical Review B}\ }\textbf {\bibinfo {volume} {90}} (\bibinfo
  {year} {2014}),\ 10.1103/physrevb.90.220103}\BibitemShut {NoStop}%
\bibitem [{\citenamefont {Bowler}\ and\ \citenamefont
  {Miyazaki}(2012)}]{Bowler2012}%
  \BibitemOpen
  \bibfield  {author} {\bibinfo {author} {\bibfnamefont {D.~R.}\ \bibnamefont
  {Bowler}}\ and\ \bibinfo {author} {\bibfnamefont {T.}~\bibnamefont
  {Miyazaki}},\ }\href {\doibase 10.1088/0034-4885/75/3/036503} {\bibfield
  {journal} {\bibinfo  {journal} {Reports on Progress in Physics}\ }\textbf
  {\bibinfo {volume} {75}},\ \bibinfo {pages} {036503} (\bibinfo {year}
  {2012})}\BibitemShut {NoStop}%
\end{thebibliography}%
\end{document}


\author{Jack S. Baker$^{1, 2}$ and David Bowler$^{1, 2}$  \\\small{$^1$\textit{London Centre for Nanotechnology, UCL, 17-19 Gordon St, London WC1H 0AH, UK}} \\ \small{$^2$\textit{Department of Physics \& Astronomy, UCL, Gower St, London WC1E 6BT, UK}}}

\title{\textbf{Supplemental Material:} \\ \medskip \normalsize{First principles soft mode lattice dynamics of PbZr$_{0.5}$Ti$_{0.5}$O$_3$ and shortcomings of the
virtual crystal approximation}}
\maketitle

\medskip


This document provides all the necessary supplemental material for the article "\textit{First principles soft mode lattice dynamics of PbZr$_{0.5}$Ti$_{0.5}$O$_3$ and shortcomings of the
virtual crystal approximation}". It is divided into three sections. Section \ref{dfptfdm:level1} demonstrates that the two methods for calculation of the phonon dispersion curves - density functional perturbation theory (DFPT) and the finite displacement method (FDM) are equivalent. This is completed by comparison of the dispersions from both methods on end members PTO \& PZO. Section \ref{fulldisp:level1} displays the full phonon dispersion relations (including the real-space) for each of the configurations treated in the original article which may be of use to some readers seeing as the articles itself deals only with dispersion in the soft-space. Section \ref{fulltab:level1} tabulates in full the soft modes of each arrangement of PZT \& end members PTO \& PZO. If further information or raw data is required, do not hesitate to contact the authors via email (jack.baker.16@ucl.ac.uk).

\section{\label{fulltab:level1} Full soft mode tabulation}

The following table shows the \textbf{analytical} phonon wavenumbers of soft modes at the considered wavevectors from the original article. This can be considered as a continuation of Table III from the original text but with all soft modes included. Modes are listed in descending order in imaginary wave number from left to right.

\begin{longtable}{@{}lccccccccc@{}}
\toprule
\multicolumn{1}{c}{}             &                                      & M & $\bar{\nu}$ {[}$i$ cm$^{-1}${]} &                                    & M  & $\bar{\nu}$ {[}$i$ cm$^{-1}${]} &                                    & M                    & $\bar{\nu}$ {[}$i$ cm$^{-1}${]} \\* \midrule
\endhead
\bottomrule
\endfoot
\endlastfoot
\multicolumn{1}{c}{\textbf{PTO}} & $\Gamma_4^-$                         & 3 & 149.60                          & $\Delta_5^+$                       & 4  & 83.40                           & R$_4^+$                            & 3                    & 62.12                           \\
\multicolumn{1}{c}{}             & T$_4$                                & 2 & 45.40                           & $\Sigma_3$                         & 4  & 37.61                           & $\Sigma_2$                         & 4                    & 27.05                           \\
\multicolumn{1}{c}{}             & M$_3^+$                              & 3 & 18.02                           &                                    &    &                                 &                                    &                      &                                 \\* \midrule
\multicolumn{1}{c}{\textbf{PZO}} & R$_4^+$                              & 3 & 181.52                          & T$_4$                              & 2  & 178.75                          & M$_3^+$                            & 3                    & 176.09                          \\
\multicolumn{1}{c}{}             & $\Gamma_4^-$                         & 3 & 132.14                          & S$_4$                              & 4  & 128.32                          & T$_5$                              & 4                    & 120.45                          \\
\multicolumn{1}{c}{}             & $\Sigma_2$                           & 4 & 119.69                          & Z$_4$                              & 4  & 119.53                          & S$_3$                              & 4                    & 105.37                          \\
\multicolumn{1}{c}{}             & $\Lambda_2$                          & 8 & 103.13                          & $\Lambda_3$                        & 16 & 101.82                          & $\Delta_5$                         & 4                    & 93.38                           \\
\bottomrule
\end{longtable}

\section{\label{dfptfdm:level1} Equivalence of DFPT and the FDM}

\begin{figure}
       \includegraphics[width=\linewidth]{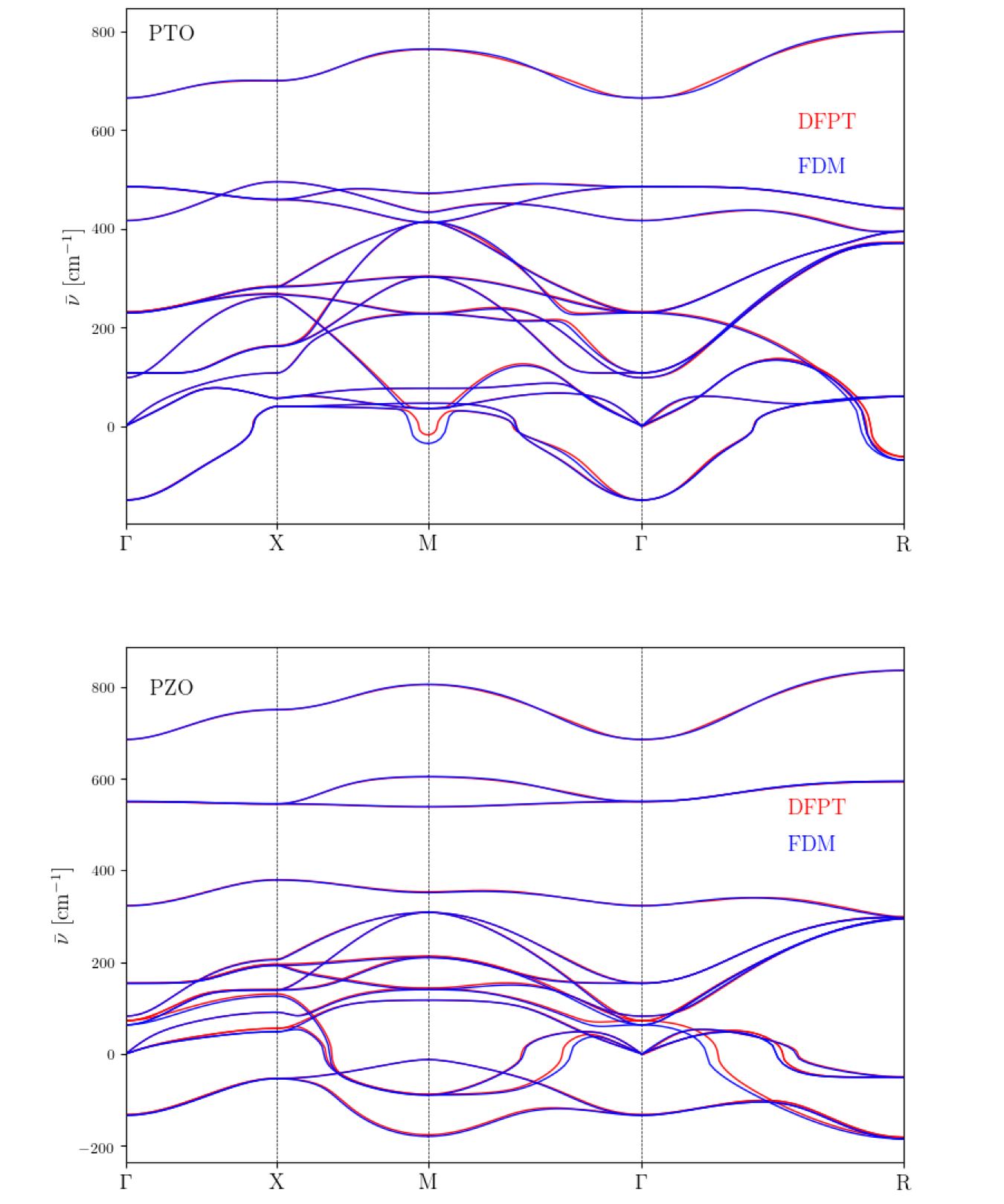}
        \caption{Phonon disprsions curves using the FDM and DFPT for both PTO (top) and PZO (bottom). This calculation is performed across the primitive cell of PTO \& PZO.}
        \label{fig:PTO_FDM_DFPT}
    \end{figure}

This section does not aim to prove the formal equivalence of the FDM and DFPT but rather demonstrates that in practicality, comparable results can be achieved between them. This test is a justification for the use of the FDM for VCA calculations in the orignal text when others are performed with DFPT. Figure \ref{fig:PTO_FDM_DFPT} shows the DFPT \& FDM dispersions of PTO \& PZO using the primitive perovskite cell. DFPT calculations are performed using the implementation in \texttt{ABINIT} (\texttt{v8.10.2}) whilst FDM calculations are performed using \texttt{phonopy} (\texttt{v2.1}). 

We use a $4\time4\times4\times4$ supercell and a displacement of 0.01$\text{\AA}$ for the FDM calculation and a $\Gamma$-centred $8 \times 8 \times 8$ $\mathbf{q}$-point mesh for the DFPT calculation. We ensure that we achieve the same level of convergence in the electronic ground state for both cases. The parameters used represent what is feasibly possible (at current) for both methods for a modest computational cost. This means that the DFPT calculation should be more accurate since the dense $\mathbf{q}$-point mesh incorporates the exact calculation of phonons at longer wavelengths than what is possible within the $4\time4\times4\times4$ supercell in the FDM. \par

Figure \ref{fig:PTO_FDM_DFPT} shows that there is clearly great agreement between the two methods across the whole of the first BZ. The calculations are however not without discrepancy. Notably, at the M-point of the PTO calculation, the soft modes of the FDM are slghtly softer than the DFPT calculation. Discrepancies of a similar magnitude also exist as we approach $\Gamma$ for the PZO calculation. These errors are as the result of interpolation on a sparser grid of $\mathbf{q}$-points for the FDM calculation vs the DFPT calculation. This level of error is however small and does not affect the general hierarchy of modes.

\section{\label{fulldisp:level1} Full phonon dispersions}

In this section we show the full phonon dispersion curves for the arrangements of PZT, PTO \& PZO referenced in the original text. The reader is reminded that these dispersions are over the fractional $\mathbf{q}$-path (0, 0, 0) $\rightarrow$ (0, 1/2, 0) $\rightarrow$ (1/2, 1/2, 0) $\rightarrow$ (1/2, 1/2, 1/2)  in some cases incurring multiple symmetry labels depending on the degree of folding in the BZ. The dispersions are displayed in figure \ref{fig:I_II} onwards.

    \begin{figure}
       \includegraphics[width=\linewidth]{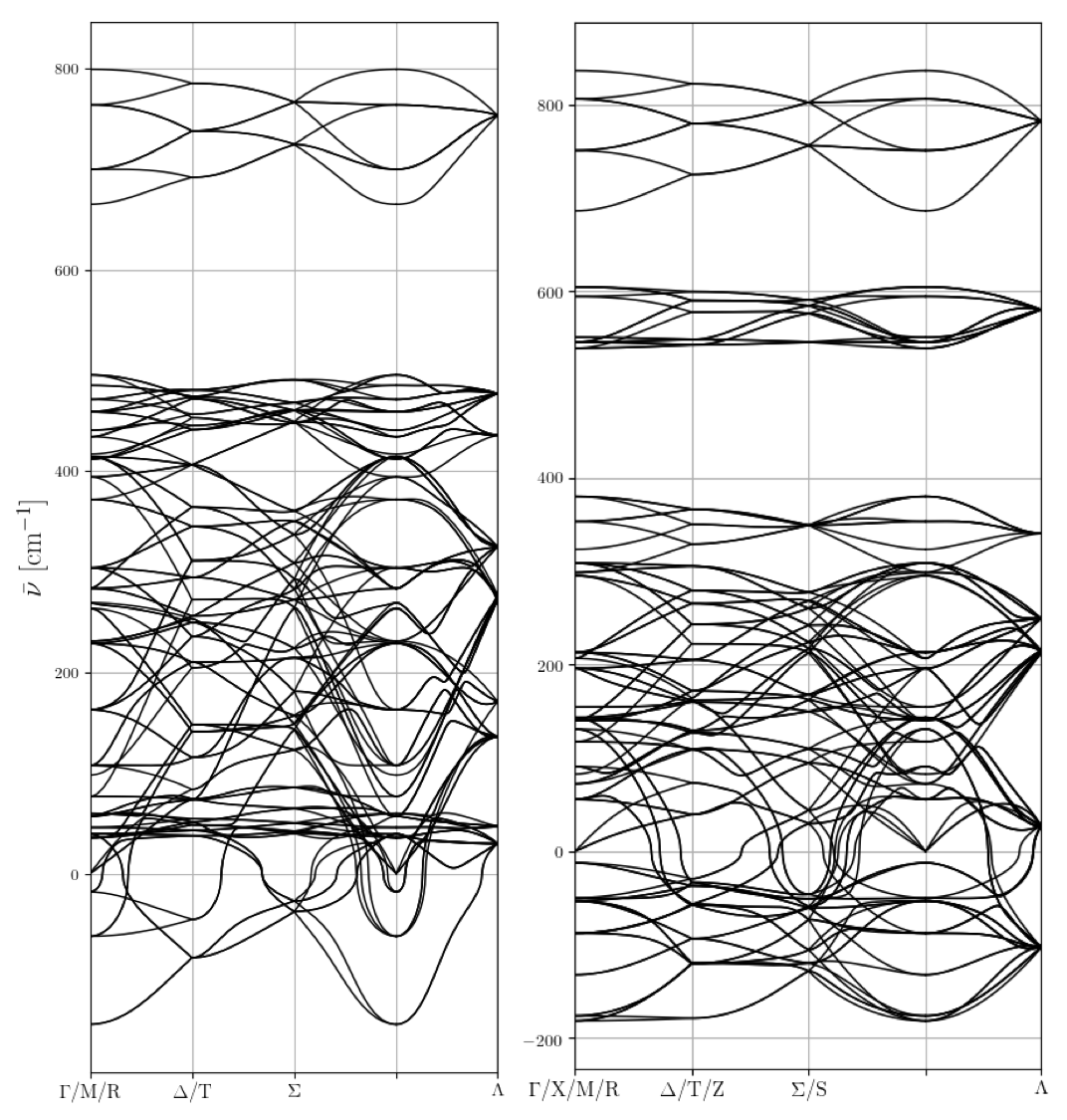}
        \caption{The full phonon dispersion curves for PTO (left) \& PZO (right) as referenced from the original article. Note that these dispersions (like most) are over a non primitive cell so will look significantly different to other dispersions of PTO \& PZO found in the literature.}
        \label{fig:I_II}
    \end{figure}

    \begin{figure}
       \includegraphics[width=\linewidth]{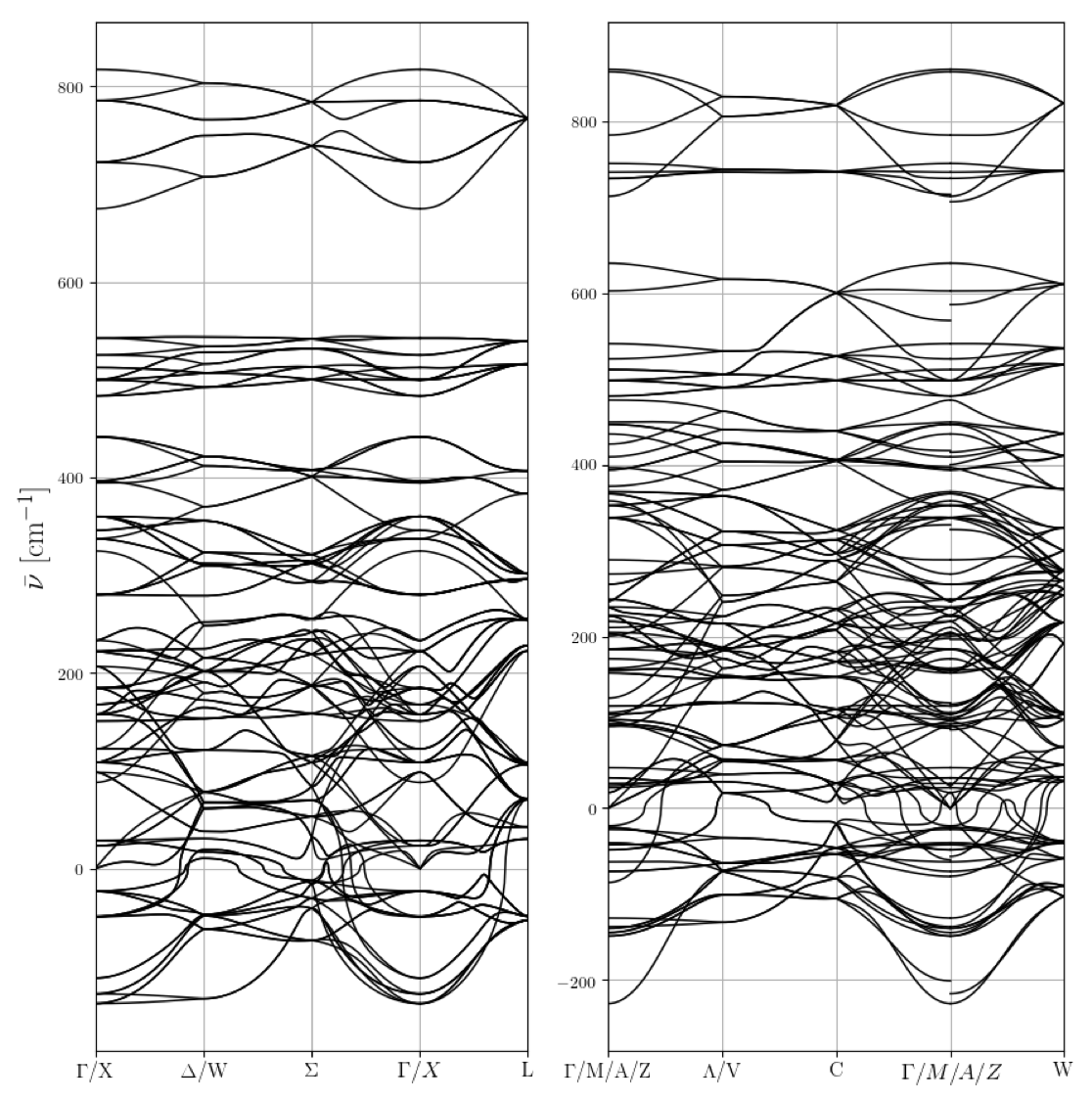}
        \caption{The full phonon dispersion curves for PZT arrangements I (left) \& II (right) as referenced from the original article.}
        \label{fig:I_II}
    \end{figure}
    
    \begin{figure}
       \includegraphics[width=\linewidth]{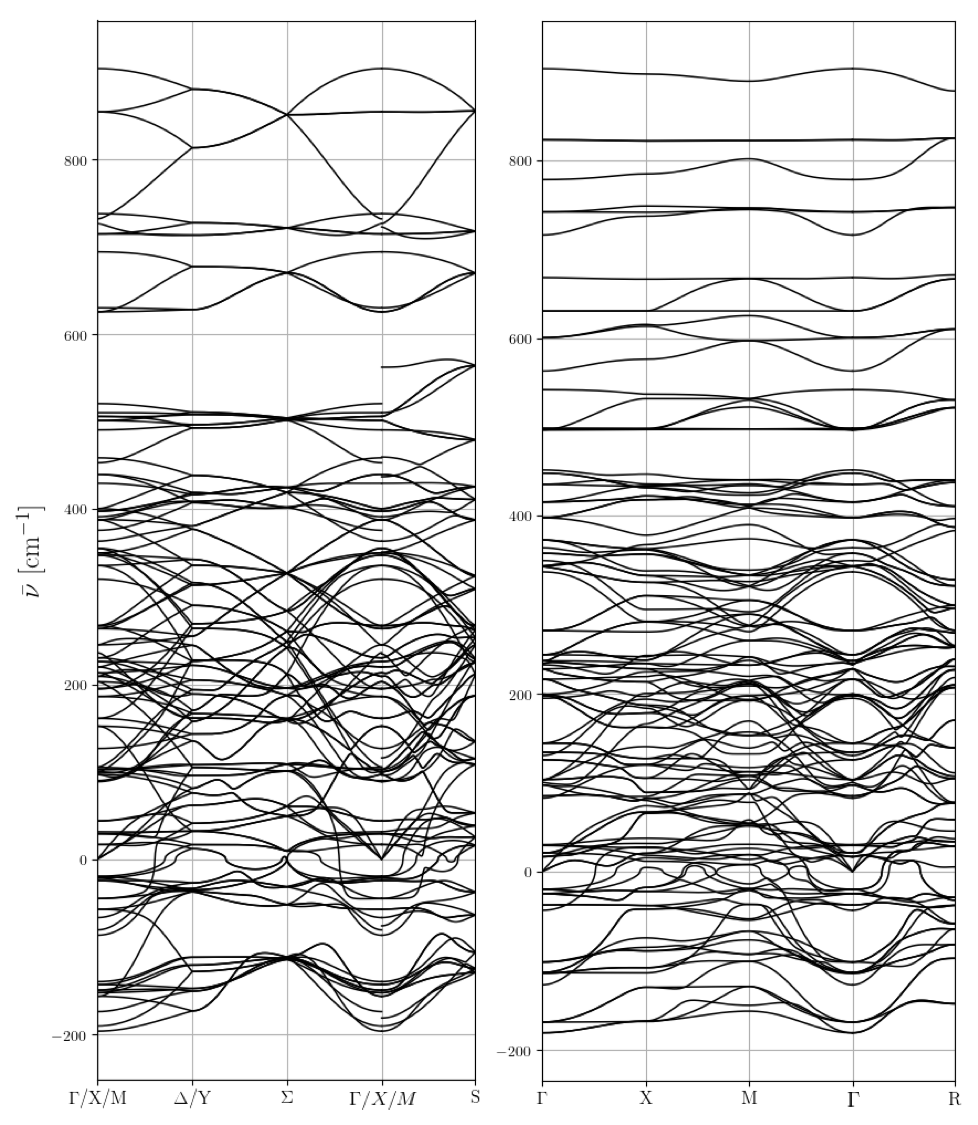}
        \caption{The full phonon dispersion curves for PZT arrangements III (left) \& IV (right) as referenced from the original article.}
        \label{fig:III_IV}
    \end{figure}
    
    \begin{figure}
       \includegraphics[width=\linewidth]{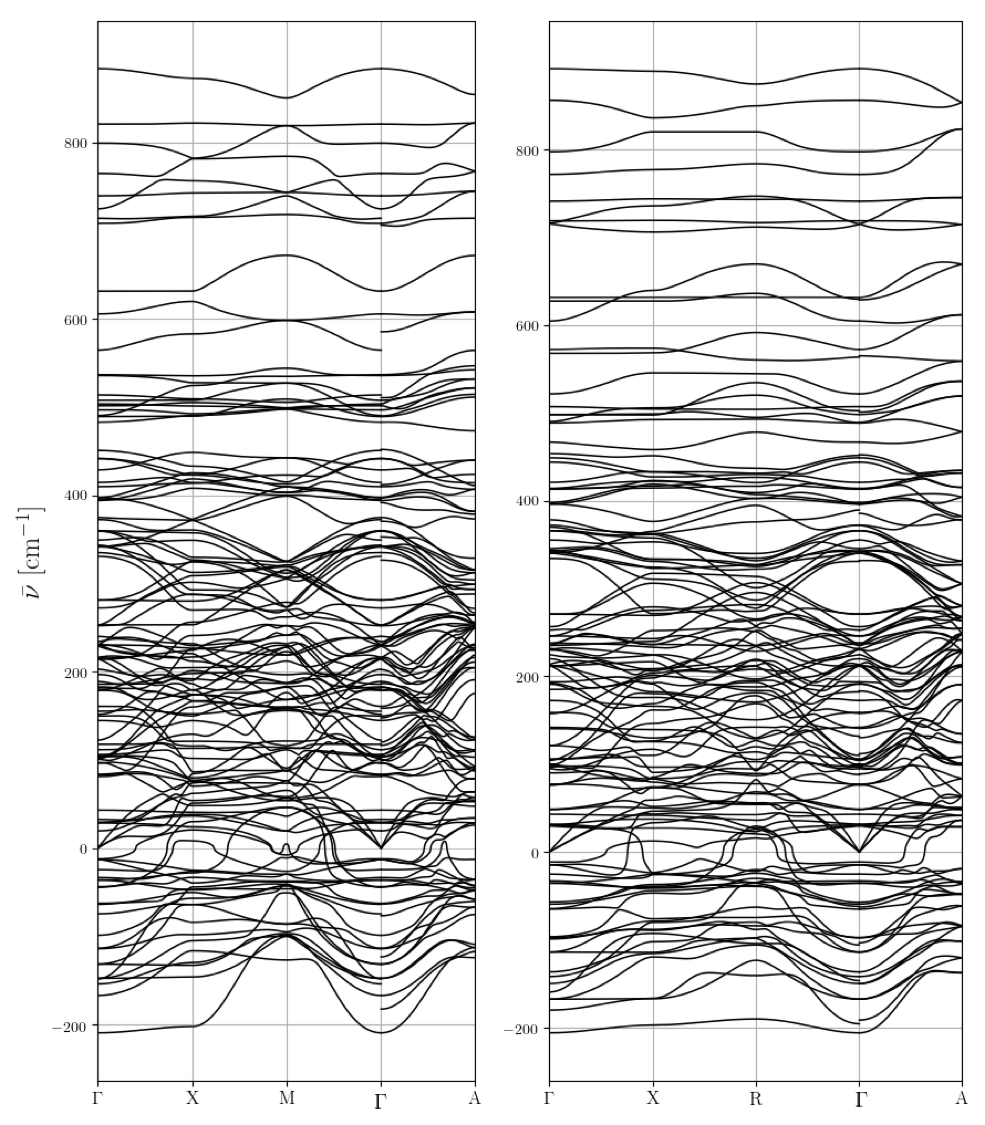}
        \caption{The full phonon dispersion curves for PZT arrangements V (left) \& VI (right) as referenced from the original article.}
        \label{fig:V_VI}
    \end{figure}
    
    \begin{figure}
    \centering
        \caption{The full phonon dispersion curves for the VCA calculation as referenced from the original article.}
       \includegraphics[width=\textwidth,height=\textheight,keepaspectratio]{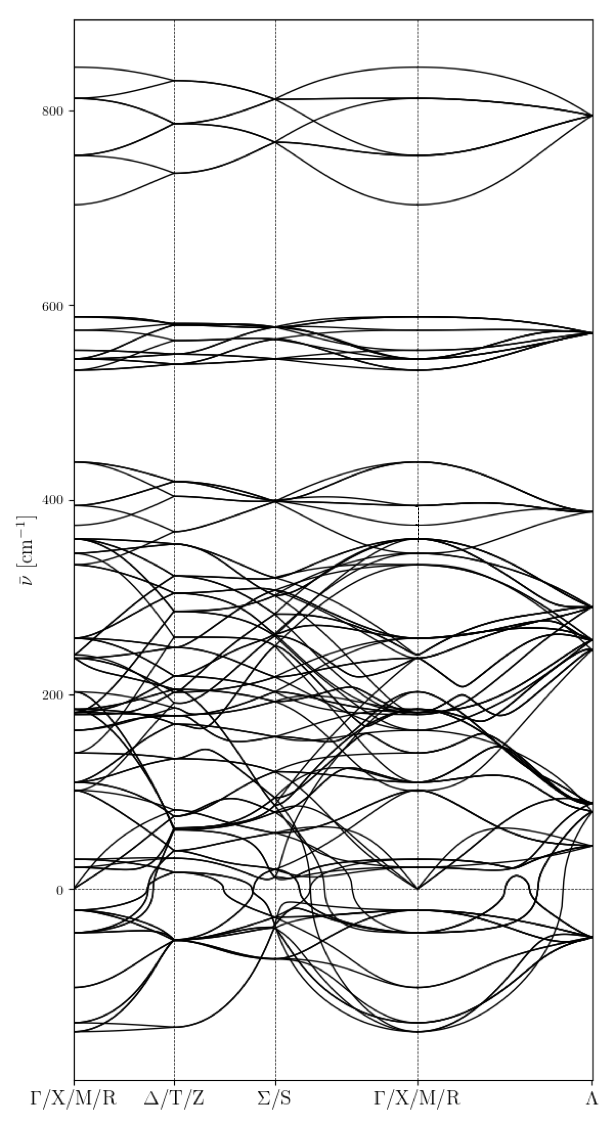}

        \label{fig:VCA}
    \end{figure}